\documentclass[a4paper,11pt]{article}

\usepackage[english]{babel}

\usepackage[a4paper,tmargin=3truecm,bmargin=3truecm,rmargin=2.5truecm,
lmargin=2.5truecm,twoside,verbose=true]{geometry}
\usepackage{cancel,graphicx}
\usepackage{hyperref}
\usepackage{amsmath,amssymb,slashed}

\usepackage{amsmath,amssymb}
\usepackage[amsmath, hyperref, thmmarks]{ntheorem}
\numberwithin{equation}{section}

\allowdisplaybreaks[1]

\newcommand\bbbone{\mathbb{I}}

\newcommand{\hd}{\hat{\mathrm{d}}}
\newcommand\caZ{{\mathcal Z}}

\newcommand\gone{{ \mathchoice {1\mskip-4mu\mathrm{l} } {1\mskip-4mu\mathrm{l} }{1\mskip-4.5mu\mathrm{l} } {1\mskip-5mu\mathrm{l}} }}

\def\gC{{\mathbb C}}

\def\ksl{{\mathfrak{sl}}}

\newcommand\algA{{\mathbf A}}

\newcommand\modM{{\boldsymbol M}}

\newcommand\kg{{\mathfrak g}}
\newcommand\hR{{F}}
\newcommand\adrep{\textup{ad}}
\newcommand\der{{\text{\textup{Der}}}}
\newcommand\kX{{X}}
\newcommand\kY{{ Y}}

\newcommand\kS{{\mathfrak S}}

\newcommand{\grast}{\bullet}

\DeclareMathOperator{\tr}{Tr}

\newcommand\Int{{\text{\textup{Int}}}}
\newcommand\out{{\text{\textup{Out}}}}
\newcommand\Out{{\text{\textup{Out}}}}
\newcommand\dd{{\text{\textup{d}}}}

\theoremsymbol{}
\theorembodyfont{\slshape}
\theoremheaderfont{\normalfont\bfseries}
\theoremseparator{}
\newtheorem{Theorem}{Theorem}[section]

\newtheorem{proposition}[Theorem]{Proposition}

\newtheorem{lemma}[Theorem]{Lemma}

\theorembodyfont{\upshape}
\theoremsymbol{\ensuremath{\blacklozenge}}

\newtheorem{definition}[Theorem]{Definition}
\theoremstyle{nonumberplain}
\theoremheaderfont{\scshape}
\theorembodyfont{\normalfont}
\theoremsymbol{\ensuremath{\blacksquare}}

\newtheorem{proof}{Proof}
\qedsymbol{\ensuremath{_\blacksquare}}
\theoremclass{LaTeX}

\renewenvironment{thebibliography}[1]
         {\section*{References}\frenchspacing\small
          \begin{list}{[\arabic{enumi}]}
         {\usecounter{enumi}\parsep=2pt\topsep 0pt
         \settowidth{\labelwidth}{[#1]}
         \leftmargin=\labelwidth\advance\leftmargin\labelsep
         \rightmargin=0pt\itemsep=1pt\sloppy}}{\end{list}}
\title{Noncommutative Yang-Mills-Higgs actions from derivation-based differential calculus\footnote{Work 
supported by ANR grant NT05-3-43374 ``GENOPHY''.}}
\author{Eric Cagnache, Thierry Masson and Jean-Christophe Wallet}

\begin{document}
\date{}
\maketitle
\vspace*{-1cm}
\begin{center}
\textit{Laboratoire de Physique Th\'eorique, B\^at.\ 210\\
    Universit\'e Paris XI,  F-91405 Orsay Cedex, France\\
    e-mail: \texttt{eric.cagnache@th.u-psud.fr, thierry.masson@th.u-psud.fr}, 
\texttt{jean-christophe.wallet@th.u-psud.fr}}\\[1ex]

\end{center}

\vskip 2cm

\begin{abstract}
Derivations of a noncommutative algebra can be used to construct differential calculi, the so-called derivation-based differential calculi. We apply this framework to a version of the Moyal algebra ${\cal{M}}$. We show that the differential calculus, generated by the maximal subalgebra of the derivation algebra of ${\cal{M}}$ that can be related to infinitesimal symplectomorphisms, gives rise to a natural construction of Yang-Mills-Higgs models on ${\cal{M}}$ and a natural interpretion of the covariant coordinates as Higgs fields. We also compare in detail the main mathematical properties characterizing the present situation to those specific of two other noncommutative geometries, namely the finite dimensional matrix algebra $M_n({\mathbb{C}})$ and the algebra of matrix valued functions $C^\infty(M)\otimes M_n({\mathbb{C}})$. The UV/IR mixing problem of the resulting Yang-Mills-Higgs models is also discussed. 
\end{abstract}

\pagebreak

\section{Introduction.}
A new family of noncommutative (NC)  field theories \cite{Douglas:2001ba}, \cite{Szabo} came under increasing scrutiny after 1998 when it was realized \cite{Schomerus}, \cite{Seiberg:1999vs} that string theory seems to have some effective regimes described by noncommutative field theories (NCFT) defined on a simple NC version of flat four dimensional space. This latter is the Moyal space which has constant commutators between space coordinates. For a mathematical description see e.g \cite{Gracia-Bondia:1987kw}, \cite{Varilly:1988jk}; for a detailed study on the relationship between the Moyal algebra and non unital extention(s) of the Connes spectral triple, see \cite{GAYRAL2004}. For reviews on noncommutative geometry, see \cite{CONNES}, \cite{CM}, \cite{LANDI}, \cite{GRACIAVAR}. However, it was noticed \cite{Minwalla:1999px,Chepelev:1999tt} that the simplest NC $\varphi^4$ model , ($\varphi$ real-valued) on the 4-dimensional Moyal space is not renormalizable due to the occurence of a phenomenon called Ultraviolet/Infrared (UV/IR) mixing  \cite{Minwalla:1999px,Chepelev:1999tt,Matusis:2000jf}. This phenomenon stems basically from the existence of nonplanar diagrams that are UV finite but nevertheless develop IR singularities which when inserted into higher order diagrams are not of the renormalizable type  \cite{Douglas:2001ba}, \cite{Szabo}. A first solution to this problem, hereafter called the "harmonic solution", was proposed in 2004 \cite{Grosse:2004yu,Grosse:2003aj}. It amounts to supplement the initial action with a simple harmonic oscillator term leading to a fully renormalisable NCFT. For recent reviews, see e.g \cite{vince}, \cite{Wallet:2007 em}. This result seems to be related to the covariance of the model under the so called Langmann-Szabo duality  \cite{Langmann:2002cc}. Other renormalisable noncommutative matter field theories have then been identified  \cite{Grosse:2003nw}, \cite{Langmann:2003if}, \cite{Langmann:2003cg}, \cite{Vignes-Tourneret:2006nb} and detailled studies of the properties of the corresponding renormalisation group flows have been carried out \cite{beta1},  \cite{Lakhoua:2007ra}, exhibiting in particular the vanishing of the $\beta$-function to all orders for the $\varphi^4_4$ model \cite{beta}.\par

But, so far, the construction of a fully renormalisable gauge theory on 4-$D$ Moyal spaces remains a challenging problem. The simplest NC analog of the Yang-Mills action given by $S_0$$=$${{1}\over{4}}\int d^Dx (F_{\mu\nu}\star F_{\mu\nu})(x)$ (in the notations used in e.g \cite{Wallet:2007 em} ), suffers from UV/IR mixing. This basically stems from the occurence of an IR singularity in the polarisation tensor $\omega_{\mu\nu}(p)$ ($p$ is some external momentum). From a standard one-loop calculation, we easily infer that
\begin{align}
\omega_{\mu\nu}(p)\sim (D-2)\Gamma({{D}\over{2}}){{{\tilde{p}}_\mu{\tilde{p}}_\nu}\over{{\pi^{D/2}(\tilde{p}}^2)^{D/2}}}+... ,\quad p\to0
\label{eq:singul1}
\end{align}
where ${\tilde{p}}_\mu$$\equiv$$\Theta_{\mu\nu}p_\nu$ and $\Gamma(z)$ denotes the Euler function. This singularity, albeit obviously transverse in the sense of the Slavnov-Taylor-Ward identities, does not correspond to some gauge invariant term. This implies that the recent alternative solution to the UV/IR mixing proposed for the NC $\varphi^4$ model in \cite{GMRT}, which roughly amounts to balance the IR singularity throught a counterterm having a similar form, cannot be extended straighforwardly (if possible at all) to the case of gauge theories.\par

Recently, the extension of the harmonic solution to the case of gauge theories has been achieved in \cite{de Goursac:2007gq} and \cite{Grosse:2007qx} (see also  \cite{deGoursac:2007qi}, \cite{Wohl2007}). These works have singled out, as potential candidate for renormalisable gauge theory on 4-$D$ Moyal space, the following generic action
\begin{align}
S=\int d^4x \Big(\frac{1}{4}F_{\mu\nu}\star F_{\mu\nu}
+\frac{\Omega^2}{4}\{{\cal{A}}_\mu,{\cal{A}}_\nu\}^2_\star
+{\kappa}{\cal{A}}_\mu\star{\cal{A}}_\mu\Big) \label{eq:decadix1}
\end{align}
in which  the 2nd and 3rd terms may be viewed as "gauge counterparts'' of the harmonic term introduced in \cite{Grosse:2004yu}. Here, ${\cal{A}}_\mu$ is the covariant coordinates, a natural gauge covariant tensorial form stemming from the existence of a canonical gauge invariant connection in the present NC framework. This action has interesting properties \cite{de Goursac:2007gq}, \cite{Grosse:2007qx} deserving further studies. For instance, gauge invariant mass terms for the gauge fields are allowed even in the absence of Higgs mechanism. Besides, the covariant coordinates appears to bear some similarity with Higgs fields. It turns out that the action \eqref{eq:decadix1} has a non-trivial vacuum, whose explicit expression has been derived very recently in \cite{vacuumym:2008}, which complicates the study of its actual renormalisability. Notice that non trivial vacuum configurations also occur within NC scalar models with harmonic term as shown in \cite{deGoursac:2007uv}. \par

In this paper, we show that most of the salient (classical) properties of NC gauge theories on the Moyal space have a natural interpretation within the framework of differential calculus based on derivations. The link between the simplest NC analog of Yang-Mills theory and some version of the spectral triple has been studied closely in \cite{GAYRAL2004}. Note that we will not examine here the viewpoint of spectral triples. Moreover, the Moyal algebra we will consider is not the one used in the construction of these spectral triples. The differential calculus based on the derivations has been settled down in \cite{DBV-1, DBV-2}, \cite{Dubois-Violette:1989vr} and \cite{DBV-3}. For an exhaustive review, see \cite{DBV-4} and references therein. This framework underlies the first prototypes of NC matrix-valued field theories \cite{Dubois-Violette:1989vq, DBV-M}, \cite{Masson:1999}. For a review, see \cite{Masson:Orsay}. Here, we consider in particular a natural modification of the minimal differential calculus  generated by the "spatial derivations", the one that implicitely underlies most of the works that appeared in the physics litterature. We show that this new differential calculus, generated by the maximal subalgebra of the derivation algebra of ${\cal{M}}$ whose elements are related to (infinitesimal) symplectomorphisms, permits one to construct NC gauge theories that can be interpreted as Yang-Mills-Higgs models on ${\cal{M}}$, the covariant coordinates of the physics litterature being interpreted as Higgs fields, thanks to the existence of a gauge invariant canonical connection. We consider models invariant under $U(1)$ or $U(n)$ gauge transformations. We also compare in detail the present situation to the other NC geometries stemming from the finite dimensional matrix algebra $M_n({\mathbb{C}})$ and for the algebra of matrix valued functions $C^\infty(M)\otimes M_n({\mathbb{C}})$. Note that a similar modification of the minimal differential calculus on ${\cal{M}}$ has been considered in \cite{italiens}. However, this work did not consider the construction of gauge theories on ${\cal{M}}$ but was only focused on the construction of subalgebras of the $D=4$ Moyal algebra from a set of constraints forming a subalgebra of the $sp(2n,{\mathbb{R}})$ algebra and the obtention of the algebra of smooth functions of ${\mathbb{R}}^3$ from a commutative limit.\par 

The paper is organised as follows. In Subsection~\ref{generalproperties}, we collect the main properties of the differential calculus based on the derivations of an associative unital $*$-algebra used in the sequel and introduce a definition of a NC connection on a module over the algebra, as a natural generalisation of ordinary connections. The specific properties and simplifications occuring when the module is equal to the algebra, which is the case relevant for (most of) the NCFT on Moyal spaces studied so far, are detailled in Subsection~\ref{caseM=A}. This provides the suitable framework to deal with models invariant under $U(1)$ gauge transformations. In Subsection~\ref{caseM=An}, we give the generalisation to the $U(n)$ case which is obtained when the module is equal to the product of $n$ copies of the algebra. In Section~\ref{gaugemoyal}, we focus on the Moyal algebra ${\cal{M}}$ whose main properties are recalled in Subsection~\ref{generalpropertiemoyal}. In Subsection~\ref{diffcalculusmoyal}, we consider the differential calculus based on the maximal subalgebra of the derivations of ${\cal{M}}$ whose elements can be related to infinitesimal symplectomorphisms. Then, a direct application of the results of Subsection~\ref{caseM=A} (resp.~\ref{caseM=An}) leads to a natural construction of Yang-Mills-Higgs models defined on ${\cal{M}}$ with $U(1)$ (resp. $U(n)$) gauge invariance. This is presented in Subsection~\ref{gaugaactionsmoyal} where we also indicate some classical properties of the NC Yang-Mills-Higgs actions obtained from the modified differential calculus. Explicit one-loop computation of the vacuum polarisation tensor, given in the appendices, shows that these latter still exhibits an IR singularity of the type given in \eqref{eq:singul1}. We also perform a comparition between these models and other gauge invariant models. In Section~\ref{discussion}, we compare the mathematical features underlying the NC differential calculus of section~\ref{gaugemoyal} to those for the NC geometry stemming from the finite dimensional matrix algebra $M_n({\mathbb{C}})$ as well as for the algebra of matrix valued functions $C^\infty(M)\otimes M_n({\mathbb{C}})$. The main point is that, in some sense, the case considered in section~\ref{gaugemoyal} interpolates between these two latter situations. \par 

\section{Derivation-based differential calculus.}
The differential calculus based on the derivations of an associative algebra has been initially introduced and developped in \cite{DBV-1,Dubois-Violette:1989vr,DBV-2} to which we refer for more details. A related review can be found in \cite{DBV-4}. This differential calculus can be viewed as a NC generalisation of the Koszul algebraic approach to differential geometry \cite{KOSZUL-T}. For other NC differential calculi, related in particular to spectral triples, see in e.g \cite{LANDI}, \cite{GRACIAVAR}. \par
Subsection~\ref{generalproperties} summarizes the main properties of the derivation-based differential calculus taken from \cite{DBV-1,Dubois-Violette:1989vr,DBV-2} that will be needed thourough the paper.  This will fix our notations and conventions.  Subsection~\ref{caseM=A} outlines the essential mathematical features underlying (most of) the studies on NC field theories defined on Moyal spaces that are sometimes overlooked or even ignored by physicists. This merely corresponds to the case where the algebra plays the role of the module used to define connections. The salient role played by the canonical connection is emphasized. Subsection~\ref{caseM=An} involves the generalisation to the case where the module is equal to the product of $n$ copies of the algebra which is the relevant for the construction of $U(n)$-invariant gauge theories.\par 

\subsection{General properties.}\label{generalproperties}
Let $\algA$ be an associative $*$-algebra with unit $\bbbone$ and center $\caZ(\algA)$. We denote the involution by $a\mapsto a^\dag$, $\forall a\in \algA$. The differential calculus based on the derivations of $\algA$ is a natural NC generalisation of the usual de Rham differential calculus on a manifold. Basically, the role of the vector fields is now played by the derivations of the algebra. In this subsection, we collect the main properties that will be used in this paper. More details can be found in \cite{DBV-1, Dubois-Violette:1989vr, DBV-2, DBV-3}. 
\begin{definition}\label{def1}
The vector space of derivations of $\algA$ is the space of linear maps defined by $\der(\algA) = \{ \kX : \algA \rightarrow \algA \ / \ \kX(ab) = \kX(a) b + a \kX(b), \forall a,b\in \algA\}$. The derivation $X\in\der(\algA)$ is called real if $(X(a))^\dag=X(a^\dag)$, $\forall a\in\algA$.
\end{definition} 
The essential properties of the spaces of derivations of $\algA$ can be summarized in the following proposition.
\begin{proposition}\label{prop1}
$\der(\algA)$ is a $\caZ(\algA)$-module for the product $(f\kX )a = f(\kX a)$, $\forall f \in \caZ(\algA)$, $\forall \kX \in \der(\algA)$ and a Lie algebra for the bracket $[\kX, \kY ]a = \kX  \kY a - \kY \kX a$, $\forall \kX,\kY \in \der(\algA)$. The vector subspace of inner derivations is defined by $\Int(\algA) = \{ \adrep_a : b \mapsto [a,b]\ / \ a \in \algA\} \subset \der(\algA)$. It is a $\caZ(\algA)$-submodule and a Lie ideal. The vector subspace of outer derivations is $\Out(\algA)=\der(\algA)/\Int(\algA)$, so that the following canonical short exact sequence of Lie algebras and $\caZ(\algA)$-modules holds:  $0\longrightarrow\Int(\algA)\longrightarrow\der(\algA)\longrightarrow\Out(\algA)\longrightarrow0$
\end{proposition}
The main features of the differential calculus based on $\der(\algA)$ are involved in the following proposition. Notice that both the Lie algebra structure and the ${\cal{Z}}(\algA)$-module structure for $\der(\algA)$ are used as essential ingredients in the construction.
\begin{proposition} \label{prop2}
Let $\underline{\Omega}^n_{\der}(\algA)$ denotes the space of $\caZ(\algA)$-multilinear antisymmetric maps from $\der(\algA)^n$ to $\algA$, with $\underline{\Omega}^0_{\der}(\algA) = \algA$ and let $\underline{\Omega}^\grast_{\der}(\algA) =\textstyle \bigoplus_{n \geq 0} \underline{\Omega}^n_{\der}(\algA)$. Then ($\underline{\Omega}^\grast_{\der}(\algA)$, $\times$, ${\hat{d}}$) is a ${\mathbb{N}}$-graded differential algebra with the product $\times$ on $\underline{\Omega}^\grast_{\der}(\algA)$ and differential ${\hat{d}}:\underline{\Omega}^n_{\der}(\algA)\to\underline{\Omega}^{n+1}_{\der}(\algA)$ satisfying ${\hat{d}}^2=0$, respectively defined for  $\forall\omega\in\underline{\Omega}^p_{\der}(\algA),\ \eta\in\underline{\Omega}^q_{\der}(\algA)$ by:
\begin{equation}
(\omega\times\eta)(\kX_1,..., \kX_{p+q}) =
 \frac{1}{p!q!} \sum_{\sigma\in \kS_{p+q}} (-1)^{{\textup{sign}}(\sigma)}
\omega(\kX_{\sigma(1)},..., \kX_{\sigma(p)}) 
\eta(\kX_{\sigma(p+1)},..., 
\kX_{\sigma(p+q)}) \label{eq:product}
\end{equation}
\begin{equation}
{\hat{d}}\omega(\kX_1,..., \kX_{p+1}) = \sum_{i=1}^{p+1} (-1)^{i+1} \kX_i \omega( \kX_1,..\vee_i.., \kX_{p+1}) \nonumber
\end{equation}
\begin{align}
 + \sum_{1\leq i < j \leq p+1} (-1)^{i+j} \omega( [\kX_i, \kX_j],..\vee_i..\vee_j.., \kX_{p+1}) \label{eq:koszul}
\end{align}
\end{proposition} 
It turns out that a differential calculus can also be built from suitable subalgebras of $\der(\algA)$. The following proposition holds:
\begin{proposition}\label{prop3}
Let $\kg \subset \der(\algA)$ denotes a Lie subalgebra which is also a $\caZ(\algA)$-submodule. Then, a restricted derivation-based differential calculus $\underline{\Omega}^\grast_\kg(\algA)$ can be built from $\kg$. It is obtained from proposition \ref{prop2} by replacing the set  $\underline{\Omega}^n_\der(\algA)$, $\forall n\in{\mathbb{N}}$, by the set  $\underline{\Omega}^n_{{\cal{G}}}$ of $\caZ(\algA)$-multilinear antisymmetric maps from $\kg^n$ to $\algA$ and still using \eqref{eq:product} and \eqref{eq:koszul}.
\end{proposition} 
In this paper, we will consider a natural NC generalisation of ordinary connections, as introduced in \cite{DBV-1, Dubois-Violette:1989vr, DBV-2} to which we refer for more details. It uses left or right finite projective modules on the associative algebra. Notice that alternative NC extensions of connections based on bimodules has been considered in \cite{DBV-3}. From now on, we denote by $\modM$ a right $\algA$-module. Let $h:\modM\otimes\modM\to\algA$ denotes a hermitean structure{\footnote{Recall that a hermitean structure is a sesquilinear map, $h:\modM\otimes\modM\to\algA$, such that $h(m_1,m_2)^\dag=h(m_2,m_1)$, $h(ma_1,ma_2)=a_1^\dag h(m1,m2)a_2$, $\forall m_1,m_2\in\modM,\ \forall a_1,a_2\in\algA$.}} on $\algA$. The connection, curvature and gauge transformations are given as follows:
\begin{definition} \label{def2}
A NC connection on $\modM$ is a linear map ${\nabla}_\kX : \modM \rightarrow \modM$ satisfying:
\begin{equation}
{\nabla}_\kX (m a) = m\kX( a) + {\nabla}_\kX (m) a,\ 
{\nabla}_{f\kX}( m) = {\nabla}_\kX (m)f,\ 
{\nabla}_{\kX + \kY} (m) = {\nabla}_\kX (m) + {\nabla}_\kY (m) \label{eq:connect-leib}
\end{equation}
$\forall \kX,\kY \in \der(\algA)$, $\forall a \in \algA$, $\forall m \in \modM$, $\forall f \in \caZ(\algA)$.  Alternatively, a NC connection is also defined by the linear map $\nabla:\modM\to\modM\otimes_\algA\underline{\Omega}^1_\der(\algA)$ such that $\nabla(ma)=\nabla(m)a+m{\hat{d}}a$, $\forall m\in\modM$, $\forall a\in\algA$ which can be further extended \cite{CONNES} to any element in $\modM\otimes_\algA \underline{\Omega}^\grast_{\der}(\algA)$. We will use both definition in the sequel.  A hermitean NC connection is a NC connection satisfying in addition $X(h(m_1,m_2))=h(\nabla_X(m_1),m_2)+h(m_1,\nabla_X(m_2))$, $\forall m_1,m_2\in\modM$, for any real derivation $X$ in $\der(\algA)$. The curvature of ${\nabla}$ is the linear map $\hR(\kX, \kY) : \modM \rightarrow \modM$ defined by
\begin{equation}
 \hR(\kX, \kY) m = [ {\nabla}_\kX,{\nabla}_\kY ] m - {\nabla}_{[\kX, \kY]}m,\ \forall\kX, \kY \in \der(\algA)
\end{equation}
\end{definition}
\begin{definition}\label{def3}
The gauge group of $\modM$ is defined as the group of automorphisms of $\modM$ as a right $\algA$-module.
\end{definition} 
\begin{proposition}\label{prop4}
For any $g$ in the gauge group of $\modM$ and for any NC connection ${\nabla}$, the map ${\nabla}^g_\kX = g^{-1}\circ {\nabla}_\kX \circ g : \modM \rightarrow \modM$ defines a NC connection. Then, one has $\hR(X,Y)^g=g^{-1}\circ \hR(X,Y) \circ g$.
\end{proposition} 
It is convenient to require that the gauge transformations are compatible with the hermitean structure, that is $h(g(m_1), g(m_2))=h(m_1,m_2)$. This defines a NC analog of unitary gauge transformations. From now on, we will only consider unitary gauge tranformations. \par 
\subsection{\texorpdfstring{The case $\modM=\algA$}{The case M=A}}\label{caseM=A}
In the special case where $\modM=\algA$, that will be the case of interest for the ensuing discussion, additional simplifications occur. It is further convenient to choose the canonical hermitean structure $h_0(a_1,a_2)=a_1^\dag a_2$.
\begin{proposition}\label{prop5}
Assume that $\modM=\algA$ and $h_0(a_1,a_2)=h(a_1,a_2)=a_1^\dag a_2$. Then:\par 
i) Any NC connection is entirely determined by $\nabla_X({\mathbb{I}})$ via $\nabla_X(a)= \nabla_X({\mathbb{I}})a+X(a)$, $\forall X\in\der(\algA)$, $\forall a\in\algA$. The 1-form connection $A\in\underline{\Omega}^1_\der(\algA)$ is defined by $A:X\to A(X)=\nabla_X({\mathbb{I}})$, $\forall X\in\der(\algA)$.\par 
ii)  A NC connection is hermitean when $\nabla_X({\mathbb{I}})^\dag=-\nabla_X({\mathbb{I}})$, for any real derivation $X$.\par 
iii) The gauge group can be identified with the group of unitary elements of $\algA$, ${\cal{U}}(\algA)$, by multiplication acting on the left of $\algA$ and one has $\nabla_X({\mathbb{I}})^g=g^\dag\nabla_X({\mathbb{I}})g+g^\dag X(g)$, $\hR(X,Y)^g=g^\dag\hR(X,Y)g$, $\forall X, Y\in\der(\algA)$, $\forall a\in\algA$.\par
\end{proposition}
\begin{proof}
i) follows directly from the definition \ref{def2} (set $m={\mathbb{I}}$ in the 1st of \eqref{eq:connect-leib}). For ii), one has $\nabla_X(a_1)^\dag a_2+a_1^\dag\nabla_X(a_2)$$=$$X(a_1^\dag a_2)+a_1^\dag(\nabla_X({\mathbb{I}})^\dag+\nabla_X({\mathbb{I}}))a_2$ where the last equality stems from the expression for $\nabla_X(a)$ given in i) and the fact that $X$ is assumed to be real. From this follows ii). For iii), use definition \ref{def3} and compatibility of gauge transformations with $h_0$ which gives $g(a)=g({\mathbb{I}})a$ and $h_0(g(a_1),g(a_2))=a_1^\dag g({\mathbb{I}})^\dag g({\mathbb{I}})a_2=h_0(a_1,a_2)$. Then, the gauge transformations for $\nabla_X({\mathbb{I}})$ and the curvature stems from proposition \ref{prop4}, the expression for $\nabla_X(a)$ in i) and the expression for $\hR(X,Y)$.
\end{proof}
\begin{definition}\label{tensorform}
A tensor 1-form is a 1-form having the following gauge transformations:
\begin{align}
 {\cal{A}}^g=g^\dag{\cal{A}}g,\ \forall g\in{\cal{U}}(\algA) \label{eq:tensortrans}
\end{align}
\end{definition}
There is a special situation where canonical gauge invariant connections can show up, as indicated in the following proposition.
\begin{proposition}\label{prop2.9}
Assume that there exists $\eta\in\underline{\Omega}^1_\der(\algA)$, such that ${\hat{d}}a=[\eta,a]$, $\forall a\in\algA$. Consider the map $\nabla^{inv}:\underline{\Omega}^0_\der(\algA)\to\underline{\Omega}^1_\der(\algA)$, $\nabla^{inv}(a)={\hat{d}}a-\eta a$, $\forall a\in\algA$, so that
$\nabla^{inv}_X(a)=X(a)-\eta(X) a$. Then, the following properties hold:\par 
i) $\nabla^{inv}$ defines a connection which is gauge invariant, called the canonical connection.\par 
ii) For any NC connection $\nabla$, ${\cal{A}}\equiv\nabla-\nabla^{inv}=A+\eta$ defines a tensor form. ${\cal{A}}(X)$, $\forall X\in\der(\algA)$ are called the covariant coordinates of $\nabla$ relative to $\eta$. 
\end{proposition}
\begin{proof}
Since any 1-form can serve as defining a connection in view of the proposition \ref{prop5}, $\nabla^{inv}(a)={\hat{d}}a-\eta a$ is a connection. Notice that it reduces to $\nabla^{inv}(a)=-a\eta$, since ${\hat{d}}a=[\eta,a]$. Then, one has $(\nabla^{inv})^g(a)=g^\dag\nabla^{inv}(ga)=g^\dag(d(ga)-\eta ga)$ $=g^\dag(-ga\eta)=-a\eta=\nabla^{inv}(a)$, which shows i). The property ii) stems simply from the definition \ref{tensorform} and the gauge transformations of a NC connection.
\end{proof}
The existence of canonical connections translates into some rather general properties of the curvatures, in particular the curvature for the canonical connection. Gauge theories defined on Moyal spaces are a particular exemple of this, as shown in the next section.
\begin{lemma} \label{lemmainnerdifferential}
Assume that there exists $\eta\in\underline{\Omega}^1_\der(\algA)$, such that ${\hat{d}}a=[\eta,a]$, $\forall a\in\algA$. Let $\hR^{inv}{(X,Y)}$ denotes the curvature for the corresponding canonical connection. Then, the following properties hold:\par 
i) $\hR^{inv}{(X,Y)}=\eta([X,Y])-[\eta(X),\eta(Y)]$ and $\hR^{inv}{(X,Y)}\in\caZ(\algA)$\par 
ii) The curvature of any NC connection defined by the tensor 1-form ${\cal{A}}$ can be written as
\begin{align}
\hR{(X,Y)}=([{\cal{A}}(X),{\cal{A}}(Y)]-{\cal{A}}{[X,Y]})-([\eta(X),\eta(Y)]-\eta([X,Y])),\  \forall X,Y\in\der(\algA) \label{eq:courbure}
\end{align}
\end{lemma}
\begin{proof}
First, from the definition of $\nabla^{inv}(a)$ in proposition \ref{prop2.9}, one infers that the 2-form curvature associated to the canonical connection is $F^{inv}(a)\equiv\nabla^{inv}(\nabla^{inv}(a))=-({\hat{d}}\eta-\eta\eta)(a)$, $\forall a\in\algA$. Then, one obtains $F^{inv}(X,Y)=\eta([X,Y])-[\eta(X),\eta(Y)]$. Then, from ${\hat{d}}a=[\eta,a]$ and ${\hat{d}}^2=0$, one has $0={\hat{d}}({\hat{d}}a)={\hat{d}}(\eta a-a\eta)=[{\hat{d}}\eta,a]-[\eta,{\hat{d}}a]=[{\hat{d}}\eta,a]-[\eta,[\eta,a]]=[{\hat{d}}\eta-\eta\eta,a]$. From this follows the second part of property i). Next, one has $\nabla_X(a)={\cal{A}}(X)a-a\eta(X)$ so that 
$[\nabla_X,\nabla_Y](a)=[{\cal{A}}(X),{\cal{A}}(Y)]a-a[\eta(X),\eta(Y)]$. Therefore $\hR{(X,Y)}(a)=([{\cal{A}}(X),{\cal{A}}(Y)]-{\cal{A}}({[X,Y])})a-a([\eta(X),\eta(Y)]-\eta({[X,Y]}))$.
\end{proof}
\subsection{\texorpdfstring{The case $\modM=\algA^n$}{The case M=An}.}\label{caseM=An}
We denote by $E_i^j$, $i,j=1,...,n$, the $n^2$ elements of the canonical basis of $M_n(\gC)$. One has $E_i^jE_k^l=\delta_k^jE_i^l$. For any unital involutive algebra $\algA$, let $\modM=\algA\times\algA\times...\times\algA\equiv \algA^{n}$ be a finite type right $\algA$-module. Let $\mu_i=(0,...,0,\bbbone,0,...0)$, $i=1,...,n$, be the $n$ elements of the canonical basis on $\modM$. For any $m\in\modM$, one has $m=\sum_{i=1}^n\mu_i\alpha^i = (\alpha^1,\alpha^2,...,\alpha^n)$ for unique $\alpha^i \in \algA$, $i=1,...,n$, and $ma=(\alpha^1 a,\alpha^2 a,...,\alpha^n a)$, $\forall m\in\modM$, $\forall a\in \algA$. Let $h_0(m_1,m_2)=m_1^\dag m_2=\sum_{i=1}^n{\alpha_{1}^{i}}^{\dag}\alpha_2^i$, $\forall m_1,m_2\in\modM$ denotes  the canonical hermitean structure. The following propositions generalise the results presented in the subsection~\ref{caseM=A}. From now on, we use the Einstein summation convention $\sum_{i=1}^nx_iy^i\equiv x_iy^i$.
\begin{proposition}\label{casun-1} For any unital involutive algebra $\algA$, any NC hermitean connection on $\modM=\algA^{n}$, $\nabla_X:\modM\to\modM$, $\forall X\in\der(\algA)$ is  fully determined by a matrix $\omega(X)\in\algA\otimes M_n(\gC)$, which is antihermitean for real $X$, defined by $\omega(X)=E_i^j\nabla_X^i(\mu_j)\equiv-iE_i^jA_j^i(X)$, where 
$\nabla_X(\mu_i)=\mu_j\nabla_X^j(\mu_i)$, $A_i^j(X)\in\algA$, $\forall i,j=1,...,n$, so that one has 
\begin{align}
\nabla_X(m)=\mu_j\nabla_X^j(\mu_i)\alpha^i+\mu_iX(\alpha^i)=\omega(X)m+X(m),\ \forall m\in\modM \label{eq:larelation11}
\end{align}
where the matrix product is understood in the second relation and $X(m)=\mu_iX(\alpha^i)$, $m=\mu_i\alpha^i$. This defines the 1-form connection $\omega\in \underline{\Omega}^1_\der(\algA)\otimes M_n(\gC)$. The 2-form curvature $F\in\underline{\Omega}^2_\der(\algA)\otimes M_n(\gC)$ is given by $Fm\equiv\nabla(\nabla(m))=({\hat{d}}\Omega+\Omega\Omega)m$, $\forall m\in\modM$ where the matrix product is again understood.
\end{proposition}
\begin{proof}
Defining $\nabla_X(\mu_i)=\mu_j\nabla^j_X(\mu_i)$, with $\nabla_X^j(\mu_i)\in \algA$, $\forall X\in\der(\algA)$, the NC connection is then determined by the matrix $\omega(X)=E_i^j\nabla_X^i(\mu_j)$, and one has immediately $\nabla_X(m)=\mu_j\nabla_X^j(\mu_i)\alpha^i+\mu_iX(\alpha^i)$. Then, by writing the matrix product $\omega(X) m=\omega(X)_i^jE^i_j\mu_k\alpha^k=\mu_j\delta^i_k\omega(X)^j_i\alpha^k=\mu_j\nabla_X^j(\mu_i)\alpha^i$, one obtains \eqref{eq:larelation11}. Finally, for any real $X$, from $X(h(m_1,m_2))=h(\nabla_X(m_1),m_2)+h(m_1,\nabla_X(m_2))$ that holds for hermitean connections, one has $m_1^\dag\omega(X)^\dag m_2+m_2^\dag\omega(X)^\dag m_1=0$ which is verified provided $\omega(X)^\dag =-\omega(X)$.
\end{proof}
The group of unitary gauge transformations ${\cal{U}}(\modM)$ is defined as the group of automorphisms of $\modM$ preserving the hermitean structure:
\begin{align}
{\cal{U}}(\modM)=\{\varphi\in{\textup{Aut}}(\modM),\ h_0(\varphi(m_1),\varphi(m_2))=h_0(m_1,m_2), \ \forall m_1, m_2\in\modM \}
\end{align}
One has the following proposition:
\begin{proposition} \label{casun-2} Assume that $\modM=\algA^{n}$:\par 
i) The unitary gauge group of $\modM$ is the group of unitary matrices $U=E_i^j\varphi^i(\mu_j)$, $U\in U(n,\algA)$ with $\varphi(\mu_i)=\mu_j\varphi^j(\mu_i)$, $\varphi^j(\mu_i)\in\algA$, $\forall \varphi\in{\cal{U}}(\modM)$, $\forall i,j=1,...,n$ and left action on $\algA\otimes M_n(\gC)$.\par 
ii) The action of the unitary gauge group ${\cal{U}}(\modM)=U(n,\algA)$ on $\omega$ is 
\begin{align}
\omega^U(X)=U^\dag\omega(X) U+U^\dag X(U),\ \forall U\in U(n,\algA),\ \forall X\in\der(\algA)
\end{align}
where the matrix product is understood.
\end{proposition}
\begin{proof}
The proof can be obtained from standard calculations. First, for any $\varphi\in{\cal{U}}(\modM)$, one has $\varphi(\mu_i)=\mu_j\varphi^j(\mu_i)$, so that $\varphi$ is fully determined by the matrix $U=E_i^j\varphi^i(\mu_j)\equiv E_j^iu^j_i$ and one has immediately $\varphi(m)=Um$ where the matrix product is understood. From the definition of unitary gauge transformations, $h_0(\varphi(m_1),\varphi(m_2))=h_0(m_1,m_2)$, $\forall m_1,m_2\in\modM$, one obtains $m_1^\dag U^\dag U m_2=m_1^\dag m_2$ which holds provided $U^\dag U=UU^\dag =\bbbone_{U(n)}$. From this follows i). To prove ii), one simply computes $\nabla^\varphi(m)=\varphi^{-1}\nabla_X(\varphi(m))$ using $\varphi(m)=Um$ and the definition of $\omega$ given in proposition \ref{casun-1}.
\end{proof}
Notice that the definition \ref{tensorform} still holds for tensor forms where now $g\in{\cal{U}}(\modM)=U(n,\algA)$ in \eqref{eq:tensortrans}. The gauge transformations of the curvature are given by $F^U=U^\dag F U$, $\forall U\in U(n,\algA)$. The proposition \ref{prop2.9} can be modified as follows.
\begin{proposition}\label{casun-3} Assume that there exists $\eta\in\underline{\Omega}^1_\der(\algA)$, such that ${\hat{d}}a=[\eta,a]$, $\forall a\in\algA$. Then, the map $\nabla_X^{inv}:\modM\to\modM$, $\nabla^{inv}_X(m)=X(m)-\eta(X)\bbbone_{U(n)}m$, $\forall X\in\der(\algA)$, $\forall m\in\modM$defines a gauge invariant connection.
\end{proposition}
\begin{proof}
It is just a straightforward computation.
\end{proof}
\section{Gauge theories on the Moyal spaces.}\label{gaugemoyal}
\subsection{General properties of the Moyal algebra.}\label{generalpropertiemoyal}
In this subsection, we collect the properties of the Moyal algebra that will be used in the sequel. For more details, see e.g \cite{Gracia-Bondia:1987kw,Varilly:1988jk, GRACIAVAR}. A recent study on the relationship between (subalgebras of) the Moyal algebra defined below and non unital extention(s) of the Connes spectral triple has been carried out in \cite{GAYRAL2004}. \par 
Let ${\cal{S}}({\mathbb{R}}^D)\equiv{\cal{S}}$ and ${\cal{S}}^\prime({\mathbb{R}}^D)\equiv{\cal{S}}^\prime$, with $D=2n$,  be respectively 
the space of complex-valued Schwartz functions on ${\mathbb{R}}^D$ and the dual space of tempered distributions on ${\mathbb{R}}^D$. The complex conjugation in ${\cal{S}}$, $a\mapsto a^\dag$, $\forall a\in{\cal{S}}$, defines a natural involution in ${\cal{S}}$ that can be extended to ${\cal{S}}^\prime$ by duality and that will be used in the rest of this paper. Let $\Theta_{\mu\nu}$ be an invertible constant skew-symmetric matrix which can be written as $\Theta=\theta{{\Sigma}}$ where $\Sigma$ is the "block-diagonal" matrix, ${{\Sigma}}=$diag$(J,...,J)$ involving $n$ $(2\times2)$ matrices $J$ given by $J =\begin{pmatrix} 0&-1 \\ 1& 0 \end{pmatrix}$ and the parameter
$\theta$ has mass dimension $-2$. We use the notation $y\Theta^{-1}z\equiv y_\mu \Theta^{-1}_{\mu\nu}z^\nu$. The following proposition summarizes properties relevant for the ensuing discussion:
\begin{proposition}
Let the $\star$-Moyal product be defined as $\star:{\cal{S}}\times{\cal{S}}\to{\cal{S}}$ by
\begin{align}
(a\star b)(x)=\frac{1}{(\pi\theta)^D}\int d^Dyd^Dz\ a(x+y)b(x+z)e^{-i2y\Theta^{-1}z},\ \forall a,b\in{\cal{S}}\label{eq:moyal}
\end{align}
Then, $({\cal{S}},\star)$ is a non unital associative involutive Fr\'echet algebra with faithfull trace given by $\int d^Dx\ (a\star b)(x)=\int d^Dx\ (b\star a)(x)=\int d^Dx\ a(x).b(x)$, where the symbol ``.'' is the usual commutative product of functions in ${\cal{S}}$.
\end{proposition}
The $\star$ product \eqref{eq:moyal} can be further extended to ${\cal{S}}^\prime\times{\cal{S}}\to{\cal{S}}^\prime$ upon using duality of vector spaces: 
$\langle T\star a,b \rangle = \langle T,a\star
b\rangle$, $\forall T\in{\cal{S}}^\prime$, $\forall a,b\in{\cal{S}}$. In a similar way, \eqref{eq:moyal} can be extended
to ${\cal{S}} \times {\cal{S}}^\prime\to{\cal{S}}^\prime$, via $\langle a\star T,b \rangle = \langle T,b\star
a\rangle$, $\forall T\in{\cal{S}}^\prime$, $\forall a,b\in{\cal{S}}$. Then, the Moyal algebra is defined as \cite{Gracia-Bondia:1987kw, Varilly:1988jk}
\begin{definition}\label{moyaldef}
Let ${\cal{M}}_L$ and ${\cal{M}}_R$ be respectively defined by ${\cal{M}}_L=\{T\in{\cal{S}}^\prime\ /\ a\star T\in{\cal{S}},\ \forall a\in{\cal{S}}\}$ and ${\cal{M}}_R=\{T\in{\cal{S}}^\prime\ /\ T\star a\in{\cal{S}},\ \forall a\in{\cal{S}}\}$. Then, the Moyal algebra ${\cal{M}}$ is defined by 
\begin{align}
{\cal{M}}={\cal{M}}_L\cap{\cal{M}}_R\label{eq:Moyal-alg}
\end{align}
\end{definition}
Notice that ${\cal{M}}_L$ and ${\cal{M}}_R$ are sometimes called in the litterature respectively the left and right multiplier algebras. By construction, ${\cal{S}}$ is a two-sided ideal of ${\cal{M}}$. The essential structural properties of the Moyal algebra that we will need are summarized in the following proposition:
\begin{proposition}
$({\cal{M}},\star)$ is a maximal unitalization of $({\cal{S}}, \star)$. It is a locally convex associative unital $*$-algebra. It involves the plane waves, the Dirac distribution and its derivatives and the polynomial functions. One has for any polynomial functions $a$, $b$ the following asymptotic formula for the $\star$-product
\begin{align}
(a\star b)(x)=a(x).b(x)+\sum_{n=1}^\infty{{1}\over{n!}}({{i}\over{2}}\Theta^{\mu_1\nu_1}{{\partial}\over{\partial x^{\mu_1}}}
{{\partial}\over{\partial y^{\nu_1}}})\cdots({{i}\over{2}}\Theta^{\mu_n\nu_n}{{\partial}\over{\partial x^{\mu_n}}}{{\partial}\over{\partial y^{\nu_n}}})a(x)b(y)\vert_{x=y} \label{eq:moyal-as}
\end{align}
\end{proposition}
\begin{proposition}\label{center-inner}
The center is ${\cal{Z}}({\cal{M}})={\mathbb{C}}$.
\end{proposition}
Other relevant properties of the $\star$-product
that hold on ${\cal{M}}$ and will be used in the sequel are 
\begin{proposition}
For any $a,b\in{\cal{M}}$, one has the following relations on ${\cal{M}}$ (we set $[a,b]_\star\equiv a\star b-b\star a$):
\begin{subequations}
\begin{align}
\partial_\mu(a\star b)=\partial_\mu a\star b+a\star\partial_\mu
b,\qquad 
(a\star b)^\dag=b^\dag\star a^\dag,\qquad 
[x_\mu,a]_\star=i\Theta_{\mu\nu}\partial_\nu a, \label{eq:relat1}
\\
x_\mu\star a=(x_\mu.a)+{{i}\over{2}}\Theta_{\mu\nu}\partial_\nu a, 
\qquad 
x_\mu(a\star b)=(x_\mu.a)\star b
-{{i}\over{2}}\Theta_{\mu\nu}a\star\partial_\nu b ,\label{eq:relat2}
\\
(x_\mu.x_\nu)\star a=x_\mu.x_\nu.a+{{i}\over{2}}(x_\mu\Theta_{\nu\beta}+x_\nu\Theta_{\mu\beta})\partial_\beta a
-{{1}\over{4}}\Theta_{\mu\alpha}\Theta_{\nu\sigma}\partial_\alpha\partial_\sigma a \label{eq:relat3}
\\
a\star(x_\mu.x_\nu)=x_\mu.x_\nu.a-{{i}\over{2}}(x_\mu\Theta_{\nu\beta}+x_\nu\Theta_{\mu\beta})\partial_\beta a
-{{1}\over{4}}\Theta_{\mu\alpha}\Theta_{\nu\sigma}\partial_\alpha\partial_\sigma a \label{eq:relat4}
\\
[(x_\mu.x_\nu.x_\rho),a]_\star=i(x_\rho x_\mu\Theta_{\nu\beta}+x_\nu x_\rho\Theta_{\mu\beta}+x_\mu x_\nu\Theta_{\rho\beta})\partial_\beta a-{{i}\over{4}}\Theta_{\mu\alpha}\Theta_{\nu\sigma}\Theta_{\rho\lambda}\partial_\alpha\partial_\sigma\partial_\lambda a) \label{eq:relat5}
\end{align}
\end{subequations}
\end{proposition}
\begin{proof}
These relations can be obtained by calculations.
\end{proof}
Notice that, as a special case of  the last relation \eqref{eq:relat1}, one obtains the celebrated relation among the "coordinate functions" defined on  ${\cal{M}}$:
\begin{align}
[x_\mu,x_\nu]_\star =i\Theta_{\mu\nu} \label{eq:comrelation}
\end{align}
Notice that \eqref{eq:comrelation} is similar to the commutation relation that generates the Heisenberg algebra, namely $[x,y]=i$. However, the algebras are different. In particular, ${\cal{M}}$ as given in definition \ref{moyaldef} cannot be generated by \eqref{eq:moyal-as} and involves in a subtle way by construction both the ordinary commutative product of functions, for which $x_\mu x_\nu=x_\nu x_\mu$, and the associative $\star$-product with which \eqref{eq:moyal-as} mimics the defining relation of Heisenberg algebra. Moreover, the topology used to define ${\cal{M}}$ relies on the theory of Schwartz distributions in an essential way, whereas there is no a priori natural topology on the (algebraically defined) Heisenberg algebra generated by \eqref{eq:comrelation}.\par
As a final remark, note that the Moyal algebra has ${\cal{Z}}({\cal{M}})={\mathbb{C}}$ as trivial center, which simplify the situation regarding all the structures of modules over ${\cal{Z}}({\cal{M}})$ that are involved in the present algebraic scheme. In the present case, these are simply replaced by vector spaces over ${\mathbb{C}}$.

\subsection{Differential calculus and inner derivations.}\label{diffcalculusmoyal}

The vector space of derivations of ${\cal{M}}$ is infinite dimensional. Then, a differential calculus based on the full derivation algebra $\der({\cal{M}})$ would give rise to gauge potentials with an infinite number of components. In view of the construction of physically oriented gauge theories on Moyal spaces, it is more convenient to deal with gauge potentials having a finite number of components. These occurs within restricted differential calculi based on finite dimensional Lie subalgebras of $\der(\algA)$, as given in proposition \ref{prop3}. In the following, we will consider two Lie subalgebras of $\der({\cal{M}})$, denoted by ${\cal{G}}_1$ and ${\cal{G}}_2$. The first one is abelian and is simply related to the "spatial derivatives" $\partial_\mu$. The resulting differential calculus underlies almost all the constructions of NCFT defined on Moyal spaces. For futher convenience, we set from now on
\begin{align}
\partial_\mu a=[i\xi_\mu,a]_\star, \quad \xi_\mu=-\Theta^{-1}_{\mu\nu}x^\nu,\quad\forall a\in {\cal{M}} \label{eq:innerxi}
\end{align}
The second derivation Lie subalgebra ${\cal{G}}_2$, such that ${\cal{G}}_1\subset{\cal{G}}_2$, is the maximal subalgebra of $\der({\cal{M}})$ whose derivations can be interpreted as infinitesimal symplectomorphisms.\par 
\begin{proposition}\label{propP2}
Let ${\cal{P}}_2\subset{\cal{M}}$ denotes the set of polynomial functions with degree $d\le2$. Let $\{a,b\}_{PB}\equiv\Theta_{\mu\nu}{{\partial a}\over{\partial x_\mu}}{{\partial b}\over{\partial x_\nu}}$ for any polynomial function $a,b\in{\cal{M}}$ denotes the Poisson bracket for the symplectic structure defined by $\Theta_{\mu\nu}$. Then, ${\cal{P}}_2$ equiped with the Moyal bracket $[,]_\star$ is a Lie algebra which verifies
\begin{align}
[P_1,P_2]_\star=i\{P_1,P_2\}_{PB},\ \forall P_1,P_2\in{\cal{P}}_2 \label{symplecto}
\end{align}
\end{proposition}
\begin{proof}
Using \eqref{eq:moyal-as}, one infers that $(P_1\star P_2)(x)$, $\forall P_1,P_2\in{\cal{P}}_2$ truncates to a finite expansion. Namely,  $(P_1\star P_2)(x)=P_1(x).P_2(x)+{{i}\over{2}}\Theta_{\mu\nu}{{\partial P_1}\over{\partial x_\mu}}{{\partial P_2}\over{\partial x_\nu}}-{{1}\over{4}}\Theta_{\mu\nu}\Theta_{\rho\sigma}{{\partial^2 P_1}\over{\partial x_\mu\partial x_\rho}}{{\partial P_2}\over{\partial x_\nu\partial x_\sigma}}$ where the last term is a constant. Then, $[P_1,P_2]_\star=i\Theta_{\mu\nu}{{\partial P_1}\over{\partial x_\mu}}{{\partial P_2}\over{\partial x_\nu}}$ from which follows the proposition.
\end{proof}
Consider now the Lie subalgebra ${\cal{G}}_2\subset\der({\cal{M}})$ which is the image of ${\cal{P}}_2$ by $\adrep$, ${\cal{G}}_2=\{X\in\der({\cal{M}}),\ /\ X=\adrep_P,\ P\in{\cal{P}}_2\}$. In order to apply the proposition \ref{prop2.9} and the lemma \ref{lemmainnerdifferential} to the present situation, one has to define properly the 1-form $\eta$ from which most of the objects entering the construction of gauge theories are derived. To do this, one defines the linear map $\eta$ as
\begin{align}
\eta: {\cal{G}}_2\to{\cal{P}}_2,\qquad /\qquad \eta(X)=P-P(0), \ \forall X\in{\cal{G}}_2 \label{eq:thetaform}
\end{align}
where $P(0)\in{\mathbb{C}}$ is the evaluation of the polynomial function $P$ at $x=0$. Then, $X(a)=\adrep_P(a)=\adrep_{\eta(X)}(a)$, $\forall X\in{\cal{G}}_2$, $\forall a\in{\cal{M}}$ and \eqref{eq:thetaform} define the 1-form $\eta$ satisfying the assumption of the proposition \ref{prop2.9}. Notice that $\eta(X)$ does not define a morphism of Lie algebras since $\eta([\partial_\mu,\partial_\nu])-[\eta(\partial_\mu),\eta(\partial_\nu)]_\star=[\xi_\mu,\xi_\nu]_\star =-i\Theta^{-1}_{\mu\nu}\ne 0$. Nevertheless, as implied by the property i) of the lemma \ref{lemmainnerdifferential}, one has $\eta([X_1,X_2])-[\eta(X_1),\eta(X_2)]\in{\mathbb{C}}$.\par
 
At this point, some remarks are in order. Proposition \ref{propP2} singles out two subalgebras of derivations, whose elements are related to infinitesimal symplectomorphisms. These are sometimes called area-preserving diffeomorphims in the physics litterature. \par 
 
The first algebra ${\cal{G}}_1$ is abelian and is simply the image by $\adrep$ of the algebra generated by the polynomials with degree 
$\le1$. It is the algebra related to the spatial derivatives $\partial_\mu$ in view of the 3rd relation of \eqref{eq:relat1} and \eqref{eq:innerxi}. One has immediately, thanks to \eqref{eq:comrelation}
\begin{align}
[\partial_\mu,\partial_\nu](a)=[\adrep_{i\xi_\mu},\adrep_{i\xi_\nu}](a)=\adrep_{[i\xi_\mu,i\xi_\nu]_\star}(a)=0, \forall a\in{\cal{M}} \label{eq:innerxi-2}
\end{align}
Note that the interpretation of $[x_\mu,a]_\star$ as a Lie derivative along a Hamiltonian vector field is obvious. The differential calculus based on ${\cal{G}}_1$ is the minimal one that can be set-up on the Moyal algebra and actually underlies most of the studies of the NCFT on Moyal spaces that appear in the litterature so far. \par 

The second algebra ${\cal{G}}_2$ is by construction the image by $\adrep$ of the algebra generated by the polynomials with degree $\le2$. It is the maximal subalgebra of $\der({\cal{M}})$ whose elements can be related to symplectomorphims. Observe that from \eqref{eq:relat3} and \eqref{eq:relat4} one has
\begin{align}
[(x_\mu.x_\nu),a]_\star=i(x_\mu\Theta_{\nu\beta}+x_\nu\Theta_{\mu\beta})\partial_\beta a\label{eq:degree2}
\end{align}
so that the bracket in the LHS can again be interpreted as the Lie derivative along a Hamiltonian vector field. Note that this is no longer true for polynomials with degree $d$$\ge$$3$, which is apparent from \eqref{eq:relat5} for $d$$=$$3$. \par 
Once ${\cal{G}}_1$ or ${\cal{G}}_2$ is choosen and the corresponding 1-form $\eta$ is determined, all the properties and mathematical status of the various objects entering the construction of gauge theories on Moyal spaces are entirely fixed from the proposition \ref{prop2.9} and the lemma \ref{lemmainnerdifferential}. The corresponding relations are summarized below for further convenience. For any $X\in{\cal{G}}_i,\ i=1,2$, one has
\begin{align}
\nabla^{inv}_X(a)=-a\star\eta(X);\quad {\cal{A}}(X)=A(X)+\eta(X) \label{eq:final1}
\end{align}
\begin{align}
 \nabla_X(a)=\nabla^{inv}_X(a)+{\cal{A}}(X)\star a=\nabla^{inv}_X(a)+(A(X)+\eta(X))\star a=X(a)+A(X)\star a\label{eq:final2}
\end{align}
\begin{align}
\hR{(X,Y)}=([{\cal{A}}(X),{\cal{A}}(Y)]-{\cal{A}}{[X,Y]})-([\eta(X),\eta(Y)]-\eta{[X,Y]})\label{eq:final3}
\end{align}
\subsection{Construction of gauge invariant actions.}\label{gaugaactionsmoyal}
\subsubsection{Minimal and extended differential calculi.}
Consider first the abelian algebra ${\cal{G}}_1$ generated by the spatial derivatives \eqref{eq:innerxi}. Then, after doing a simple rescaling $A(X)\to-iA(X)$ (i.e defining $\nabla_X({\mathbb{I}})\equiv-iA(X)$ so that hermitean connections verify $A(X)^\dag=A(X)$ for any real derivation $X$) in order to make contact with the notations of e.g \cite{de Goursac:2007gq,Wallet:2007 em}, and defining $\eta(\partial_\mu)\equiv\eta_\mu$, ${\cal{A}}(\partial_\mu)\equiv {\cal{A}}_\mu$, $A(\partial_\mu)\equiv A_\mu$, $F(\partial_\mu,\partial_\nu)\equiv F_{\mu\nu}$, $\mu=1,...,D$, a straighforward application of \eqref{eq:thetaform}, and \eqref{eq:final1}-\eqref{eq:final3} yields
\begin{align}
\eta_\mu=i\xi_\mu;\quad \nabla^{inv}_\mu(a)=-ia\star\xi_\mu,\ \forall a\in{\cal{M}}
\end{align}
\begin{align}
{\cal{A}}_\mu=-i(A_\mu-\xi_\mu);\quad \nabla_\mu(a)=-ia\star\xi_\mu+{\cal{A}}_\mu\star a=\partial_\mu a-iA_\mu\star a,\  \forall a\in{\cal{M}} \label{eq:tensorforma}
\end{align}
\begin{align}
F_{\mu\nu}=[{\cal{A}}_\mu,{\cal{A}}_\nu]_\star-i\Theta^{-1}_{\mu\nu}=-i\big(\partial_\mu A_\nu-\partial_\nu A_\mu-i[A_\mu,A_\nu]_\star\big) \label{eq:curvature-simple}
\end{align}
which fixe the respective mathematical status of the objects involved in most of the studies of NCFT on Moyal spaces. The group of (unitary) gauge transformation is the group of unitary elements of ${\cal{M}}$, ${\cal{U}}({\cal{M}})$, as defined in Section~\ref{generalproperties} and one has
\begin{align}
A_\mu^g=g\star A_\mu\star g^\dag+ig\star \partial_\mu g^\dag;\ {\cal{A}}_\mu^g=g\star {\cal{A}}_\mu\star g^\dag;\ F_{\mu\nu}^g=g\star F_{\mu\nu}\star g^\dag,\ \forall g\in{\cal{U}}({\cal{M}})
\end{align}
Consider now the algebra ${\cal{G}}_2$. Let ${\bar{{\cal{G}}}}_2\subset {\cal{G}}_2$ denotes the subspace of ${\cal{G}}_2$ whose image in ${\cal{M}}$ by the map $\eta$ \eqref{eq:thetaform} corresponds to the monomials of degree $2$. The image involves ${{D(D+1)}\over{2}}$ elements $X_{(\mu\nu)}\in{\bar{{\cal{G}}}}_2$ defined by
\begin{align}
\eta(X_{(\mu\nu)})=i\xi_\mu\xi_\nu\equiv\eta_{(\mu\nu)},\ \forall\mu,\nu=1,...,D
\end{align}
where the symbol $(\mu\nu)$ denotes symmetry under the exchange $\mu\leftrightarrow\nu$. Notice that the definition for the $X_{(\mu\nu)}$'s corresponds to real derivations. One has
\begin{align}
[\eta_{(\mu\nu)},\eta_{(\rho\sigma)}]_\star=-(\Theta^{-1}_{\rho\nu}{\eta_{(\mu\sigma)}}+\Theta^{-1}_{\sigma\nu}{\eta_{(\mu\rho)}}
+\Theta^{-1}_{\rho\mu}{\eta_{(\nu\sigma)}}+\Theta^{-1}_{\sigma\mu}{\eta_{(\nu\rho)}})\label{eq:slnr}
\end{align}
which define the generic commutation relations for the $sp(2n,{\mathbb{R}})$ algebra so that ${\bar{{\cal{G}}}}_2$ is a Lie subalgebra of ${\cal{G}}_2$. Then, the algebra ${\cal{G}}_2$ we choose is generated by $\{\partial_\mu, X_{(\mu\nu)}\}$. It is the algebra $isp(2n,{\mathbb{R}})$. One has the additional commutation relations
\begin{align}
[\eta_\mu,\eta_{(\rho\sigma)}]_\star=(\Theta^{-1}_{\mu\rho}\eta_\sigma+\Theta^{-1}_{\mu\sigma}\eta_\rho)\label{eq:addicom-1}
\end{align}
Notice that any derivation related to $isp(2n,{\mathbb{R}})$ can be viewed as generating an infinitesimal symplectomorphism, as discussed above. Accordingly, the subalgebra ${\cal{G}}_1\subset{\cal{G}}_2$ can actually be interpreted physically as corresponding to spatial directions while ${\bar{{\cal{G}}}}_2$ corresponds to (symplectic) rotations. 
Notice also that in the case $D=2$, upon defining
\begin{align}
\eta({X_1})={{i}\over{4{\sqrt{2}}\theta}}(x_1^2+x_2^2),\quad \eta({X_2})={{i}\over{4{\sqrt{2}}\theta}}(x_1^2-x_2^2),\quad
\eta({X_3})={{i}\over{2{\sqrt{2}}\theta}}(x_1x_2) \label{eq:sp2r}
\end{align}
one would obtain the following commutation relations 
\begin{align}
[\eta({X_1}),\eta({X_2})]_\star={{1}\over{{\sqrt{2}}}}\eta({X_3}),\quad[\eta({X_2}),\eta({X_3})]_\star=-{{1}\over{{\sqrt{2}}}}\eta({X_1}),\quad
[\eta({X_3}),\eta({X_1})]_\star={{1}\over{{\sqrt{2}}}}\eta({X_2}) \label{eq:isp2r}
\end{align}
\begin{subequations}
\begin{align}
[\eta_1,\eta({X_1})]_\star={{1}\over{2{\sqrt{2}}}}\eta_2\ ,\ [\eta_2,\eta({X_1})]_\star=-{{1}\over{2{\sqrt{2}}}}\eta_1
\end{align}
\begin{align}
[\eta_1,\eta({X_2})]_\star={{1}\over{2{\sqrt{2}}}}\eta_2\ ,\ [\eta_2,\eta({X_2})]_\star={{1}\over{2{\sqrt{2}}}}\eta_1
\end{align}
\begin{align}
[\eta_1,\eta({X_3})]_\star=-{{1}\over{2{\sqrt{2}}}}\eta_1\ ,\ [\eta_2,\eta({X_3})]_\star={{1}\over{2{\sqrt{2}}}}\eta_2
\end{align}
\end{subequations}
therefore making contact with the work carried out in \cite{italiens}. Note that \cite{italiens} did not consider the construction of gauge theories on Moyal spaces but was only focused on the construction of subalgebras of the $D=4$ Moyal algebra starting from a set of constraints forming a subalgebra of the $sp(2n,{\mathbb{R}})$ algebra and the obtention of the algebra of smooth functions of ${\mathbb{R}}^3$ from a commutative limit.\par  
A direct application of \eqref{eq:thetaform}, and \eqref{eq:final1}-\eqref{eq:final3} permits one to determine the invariant connection and the tensor form. One obtains
\begin{align}
\nabla^{inv}_{\partial_\mu}(a)\equiv\nabla^{inv}_\mu(a)=-ia\star\xi_\mu,\quad\nabla^{inv}_{X_{(\mu\nu)}}(a)\equiv \nabla^{inv}_{(\mu\nu)}(a)=-ia\star(\xi_\mu\xi_\nu)
\end{align} 
\begin{align}
{\cal{A}}(\partial_\mu)\equiv{\cal{A}}_\mu=-i(A_\mu-\xi_\mu),\quad {\cal{A}}(X_{(\mu\nu)})\equiv{\cal{A}}_{(\mu\nu)}=-i(A_{(\mu\nu)}-\xi_\mu\xi_\nu) \label{arondfinal}
\end{align} 
Then, any NC connection is obtained as the sum of the canonical connection and the tensor form, namely  
\begin{align}
\nabla_\mu(a)=\nabla^{inv}_\mu(a)+{\cal{A}}_\mu\star a=\partial_\mu a-iA_\mu\star a
\end{align}
\begin{align}
\nabla_{(\mu\nu)}(a)=\nabla^{inv}_{(\mu\nu)}(a)+{\cal{A}}_{(\mu\nu)}\star a=[i\xi_\mu\xi_\nu,a]_\star-iA_{(\mu\nu)}\star a
\end{align}

From this, one obtains the following expressions for the curvature
\begin{proposition}\label{curvature37}
Consider the differential calculus based on the maximal subalgebra of derivations of the Moyal algebra related to symplectomorphisms. 
The components of the 2-form curvature of a NC connection defined by a tensor $1$-form with components ${\cal{A}}_\mu,{\cal{A}}_{(\mu\nu)}$ are given by
\begin{align}
F(\partial_\mu,\partial_\nu)\equiv F_{\mu\nu}=[{\cal{A}}_\mu,{\cal{A}}_\nu]_\star-i\Theta^{-1}_{\mu\nu} \label{curv1}
\end{align} 
\begin{align}
F(\partial_\mu ,X_{(\rho\sigma)})\equiv F_{\mu(\rho\sigma)}=[{\cal{A}}_\mu,{\cal{A}}_{(\rho\sigma)}]_\star-\Theta^{-1}_{\mu\rho}{\cal{A}}_\sigma -\Theta^{-1}_{\mu\sigma}{\cal{A}}_\rho \label{curv2-1}
\end{align} 
\begin{align}
F(X_{(\mu\nu)},X_{(\rho\sigma)})\equiv F_{(\mu\nu)(\rho\sigma)}=[{\cal{A}}_{(\mu\nu)},{\cal{A}}_{(\rho\sigma)}]_\star+\Theta^{-1}_{\rho\nu}{\cal{A}}_{(\mu\sigma)} + \Theta^{-1}_{\sigma\nu}{\cal{A}}_{(\mu\rho)}+\Theta^{-1}_{\rho\mu}{\cal{A}}_{(\nu\sigma)} +\Theta^{-1}_{\sigma\mu}{\cal{A}}_{(\nu\rho)}\label{curv3-1}
\end{align} 
\end{proposition}
\begin{proof}
Use $\hR^{inv}{(X,Y)}\equiv\eta{[X,Y]}-[\eta(X),\eta(Y)]$ to evaluate the curvature for the canonical connection. Consider first $F^{inv}_{\mu\nu}$. From linearity of $\eta$, $[\partial_\mu,\partial_\nu]=0$ and $[\eta_\mu,\eta_\nu]=i\Theta^{-1}_{\mu\nu}$, one finds $F^{inv}_{\mu\nu}=-i\Theta^{-1}_{\mu\nu}$. Then, from \eqref{eq:final3}, one gets \eqref{curv1}. To obtain \eqref{curv2-1}, compute $[\partial_\mu,X_{(\rho\sigma)}](a)=[\adrep_{\eta_\mu},\adrep_{\eta_{(\rho\sigma)}}]=\adrep_{[\eta_\mu,\eta_{(\rho\sigma)}]_\star}$ using \eqref{eq:addicom-1}. This yields $[\partial_\mu,X_{(\rho\sigma)}](a)=\Theta^{-1}_{\mu\rho}\partial_\sigma a+\Theta^{-1}_{\mu\sigma}\partial_\rho a$ so that $\eta([\partial_\mu,X_{(\rho\sigma)}])=\Theta^{-1}_{\mu\rho}\eta_\sigma +\Theta^{-1}_{\mu\sigma}\eta_\rho$ which yields $F^{inv}_{\mu(\rho\sigma)}=0$. This combined with \eqref{eq:final3} yields \eqref{curv2-1}. For \eqref{curv3-1}, compute $[X_{(\mu\nu)},X_{(\rho\sigma)}](a)=\adrep_{[\eta_{(\mu\nu)},\eta_{(\rho\sigma)}]_\star}$ using \eqref{eq:slnr}. A straighforward calculation yields $F^{inv}_{(\mu\nu)(\rho\sigma)}=0$. From this follows \eqref{curv3-1}.
\end{proof}
The gauge transformations are
\begin{align}
A_\mu^g=g^\dag\star A_\mu\star g+ig^\dag\star\partial_\mu g,\quad A_{(\mu\nu)}^g=g^\dag\star A_{(\mu\nu)}\star g+ig^\dag\star(\xi_\mu\partial_\nu+\xi_\nu\partial_\mu)g,\ \forall g\in{\cal{U}}({\cal{M}})
\end{align}
\begin{align}
{\cal{A}}_\mu^g=g^\dag\star{\cal{A}}_\mu\star g,\quad {\cal{A}}_{(\mu\nu)}^g=g^\dag\star{\cal{A}}_{(\mu\nu)}\star g,\ \ \forall g\in{\cal{U}}({\cal{M}})
\end{align}
\begin{align}
F_{\mu\nu}^g=g^\dag\star F_{\mu\nu}\star g,\ F_{\mu(\rho\sigma)}^g=g^\dag\star F_{\mu(\rho\sigma)}\star g,\   F_{(\mu\nu)(\rho\sigma)}^g=g^\dag\star F_{(\mu\nu)(\rho\sigma)}\star g,\ \forall g\in{\cal{U}}({\cal{M}})
\end{align}
\subsubsection{Gauge invariant actions.}
A possible construction of a NC gauge theory defined from the curvature \eqref{curv1}-\eqref{curv3} can be done as follows. Let $[x]$ denotes the mass dimension{\footnote{We work in the units $\hbar=c=1$.}} of the quantity $x$. First, perform the rescaling $\eta_{(\mu\nu)}\to\mu\theta\eta_{(\mu\nu)}$ where $\mu$ is a parameter (not to be confused with a $\mu$ indice) with $[\mu]=1$ that will fixe the mass scale of the Higgs field to be identified in a while. Accordingly, the commutation relations are modified as
\begin{align}
[\eta_{(\mu\nu)},\eta_{(\rho\sigma)}]_\star=-(\mu\theta)^{-1}(\Theta^{-1}_{\rho\nu}{\eta_{(\mu\sigma)}}+\Theta^{-1}_{\sigma\nu}{\eta_{(\mu\rho)}}
+\Theta^{-1}_{\rho\mu}{\eta_{(\nu\sigma)}}+\Theta^{-1}_{\sigma\mu}{\eta_{(\nu\rho)}}) \nonumber
\end{align}
\begin{align}
[\eta_\mu,\eta_{(\rho\sigma)}]_\star=(\mu\theta)^{-1}(\Theta^{-1}_{\mu\rho}\eta_\sigma+\Theta^{-1}_{\mu\sigma}\eta_\rho)\label{eq:addicom}
\end{align}
and the components of the curvature becomes
\begin{align}
 F_{\mu(\rho\sigma)}=[{\cal{A}}_\mu,{\cal{A}}_{(\rho\sigma)}]_\star-\mu\theta(\Theta^{-1}_{\mu\rho}{\cal{A}}_\sigma 
+\Theta^{-1}_{\mu\sigma}{\cal{A}}_\rho) \label{curv2}
\end{align} 
\begin{align} F_{(\mu\nu)(\rho\sigma)}=[{\cal{A}}_{(\mu\nu)},{\cal{A}}_{(\rho\sigma)}]_\star+\mu\theta(\Theta^{-1}_{\rho\nu}{\cal{A}}_{(\mu\sigma)} + \Theta^{-1}_{\sigma\nu}{\cal{A}}_{(\mu\rho)}+\Theta^{-1}_{\rho\mu}{\cal{A}}_{(\nu\sigma)} +\Theta^{-1}_{\sigma\mu}{\cal{A}}_{(\nu\rho)})\label{curv3}
\end{align} 
with \eqref{curv1} unchanged. Next, introduce a dimensionfull coupling constant $\alpha$ with mass dimension $[\alpha]=2-n$ ($D=2n$).\par 
The  $U(1,{\cal{M}})$-invariant action is then defined by
\begin{align}
S_{{\cal{G}}_2}(A_\mu, {\cal{A}}_{(\mu\nu)})=-{{1}\over{\alpha^2}}\int d^{2n}x\big(F_{\mu\nu}\star F_{\mu\nu}+F_{\mu(\rho\sigma)}\star F_{\mu(\rho\sigma)}+ F_{(\mu\nu)(\rho\sigma)}\star F_{(\mu\nu)(\rho\sigma)}\big)\label{eq:actionym}
\end{align}
and is choosen to depend on the fields $A_\mu$ and ${\cal{A}}_{(\mu\nu)}$. The mass dimensions are $[A_\mu]=[{\cal{A}}_\mu]=[{\cal{A}}_{(\mu\nu)}]=1$.\par 
Several comments are now in order: \par 
i) The purely spatial part \eqref{curv1}  takes the expected form when expressed in term of $A_\mu$ through the 1st relation \eqref{arondfinal}, namely one obtains easily  $F_{\mu\nu}=-i(\partial_\mu A_\nu-\partial_\nu A_\mu-i[A_\mu,A_\nu]_\star)$. \par 

ii) One observes that $F_{\mu(\rho\sigma)}$ can be reexpressed as
\begin{align}
F_{\mu(\rho\sigma)}=D^A_\mu{\cal{A}}_{(\rho\sigma)}-\mu\theta(\Theta^{-1}_{\mu\rho}{\cal{A}}_\sigma 
+\Theta^{-1}_{\mu\sigma}{\cal{A}}_\rho), \quad 
D^A_\mu{\cal{A}}_{(\rho\sigma)}\equiv\partial_\mu{\cal{A}}_{(\rho\sigma)}-i[A_\mu,{\cal{A}}_{(\rho\sigma)}]_\star
\end{align}
using \eqref{eq:innerxi} and \eqref{arondfinal}. $D^A_\mu{\cal{A}}_{(\rho\sigma)}$ can be interpreted as a "covariant derivative" describing a NC analog of the minimal coupling to the covariant field ${\cal{A}}_{(\mu\nu)}$. Besides, one has
\begin{align}
-{{1}\over{\alpha^2}}\int d^{2n}x F_{\mu(\rho\sigma)}\star F_{\mu(\rho\sigma)}=-{{1}\over{\alpha^2}}\int d^{2n}x (D^A_\mu{\cal{A}}_{(\rho\sigma)})^2 \nonumber\\
-4\mu\theta(D^A_\mu{\cal{A}}_{(\rho\sigma)})\Theta^{-1}_{\mu\sigma}{\cal{A}}_\sigma +(4n+2)\mu^2{\cal{A}}_\mu{\cal{A}}_\mu
\label{eq:2emeterm}
\end{align}
The fact that $D^A_\mu{\cal{A}}_{(\rho\sigma)}$ in \eqref{eq:actionym} can be viewed as a NC analog of the covariant derivative of ${{D(D+1)}\over{2}}$ scalar fields ${\cal{A}}_{(\mu\nu)}$, is very reminiscent of a Yang-Mills-Higgs action for which the covariant coordinates ${\cal{A}}_{(\mu\nu)}$ play the role of Higgs fields. Then, the last term in the action \eqref{eq:actionym} which is the square of $F_{(\mu\nu)(\rho\sigma)}$ can be interpreted as the Higgs (quartic) potential part. Therefore, the use of a differential calculus based on the maximal subalgebra of $\der({\cal{M}})$ whose elements generate infinitesimal symplectomorphisms permits one to construct naturally NC analogs of Yang-Mills-Higgs actions defined on Moyal space.\par 

iii) In view of the last term in \eqref{eq:2emeterm}, the gauge invariant action \eqref{eq:actionym} involves a mass term for the gauge potential proportional to $\sim {{(4n+2)\mu^2}\over{\alpha}}{{A}}_\mu{{A}}_\mu$. Therefore, bare mass terms for $A_\mu$ can appears while preserving the gauge invariance of the action. Notice that the translational invariance of the action is broken by the term $(4n+2)\mu^2{\cal{A}}_\mu{\cal{A}}_\mu$ in view of ${\cal{A}}_\mu=-i(A_\mu-\xi_\mu)$. \par 

Yang-Mills-Higgs models defined on Moyal spaces can then be obtained from actions built from the square of the curvature \eqref{curv1}-\eqref{curv3} within the differential calculus based on the subalgebra ${\cal{G}}_2\subset\der({\cal{M}})$. Each additional inner derivation supplementing the "ordinary spatial derivations", which may be viewed as related to an ``extra noncommutative dimension'', corresponds to an additional covariant coordinate that can be interpreted as a Higgs field. Covariant coordinates have thus a natural interpretation as Higgs fields within the framework of the present derivation-based differential calculus. In this respect, these Yang-Mills-Higgs type actions share common features with the actions obtained from the derivation-based differential calculus first considered in \cite{Dubois-Violette:1989vq} for $\algA=C^\infty(M)\otimes M_n(\gC)$ where $M$ is a smooth finite dimensional manifold and $M_n(\gC)$ is the finite dimensional algebra of $n\times n$ matrices. This will be developped below and in section~\ref{discussion}.\par 

The extension of the present construction to $U(n)$ gauge invariant models on Moyal spaces is obtained from a straightforward application of the material given in the subsection~\ref{caseM=An}, once a derivation-based differential calculus is chosen. It is convenient to set $\omega(X)=-iA(X)=-iA^A(X)T^A$, $\forall X\in{\cal{G}}$, where $T^A$, $A=0,1,...,n^2-1$ denote the generators of the ${\mathfrak{u}}(n)$ algebra. The corresponding relevant formulas are given in Appendix~\ref{uncasepolar}. \par 
In the case of the minimal differential calculus where ${\cal{G}}={\cal{G}}_1$ is the abelian algebra generated by the spatial derivations \eqref{eq:innerxi}, one obtains immediately the expression of the curvature $F(\partial_\mu,\partial_\nu)\equiv F_{\mu\nu}$ given by
\begin{align}
F_{\mu\nu}=F^C_{\mu\nu}T^C=-i(\partial_\mu A_\nu-\partial_\nu A_\mu-i[A_\mu,A_\nu]_\star)= \nonumber
\\
-i(\partial_\mu A^C_\nu-\partial_\nu A^C_\mu+{{1}\over{2}}f^{ABC}\{A^A_\mu,A^B_\nu\}_\star-{{i}\over{2}}d^{ABC}[A^A_\mu,A^B_\nu]_\star)T^C
\end{align}
where the symbols $f^{ABC}$ and $d^{ABC}$ have been defined in Appendix~\ref{uncasepolar}. The gauge invariant action is defined by
\begin{align}
S_{{\cal{G}}_1}^\prime=-{{1}\over{\alpha^2}}\tr_{{\mathfrak{u}}(n)}\int d^{2n}xF_{\mu\nu}\star F_{\mu\nu} \label{eq:unaction}
\end{align}
where the parameter $\alpha$ has been introduced at the beginning of this section. \par
In the case of the extended differential calculus generated by ${\cal{G}}_2$, a straighforward application of the above framework yields the expressions for the components of the curvature. One has the following proposition:
\begin{proposition}
Consider the extended differential calculus generated by ${\cal{G}}_2$ on the Moyal algebra ${\cal{M}}$. For any NC hermitean connection on $\modM={\cal{M}}^n$ defined by the tensor form with components ${\tilde{{\cal{A}}}}_\mu={\cal{A}}^A_\mu T^A$, ${\tilde{{\cal{A}}}}_{(\mu\nu)}={\cal{A}}^A_{(\mu\nu)}T^A$ where $T^A$, $A=0,1,...,n^2$ are the generators of ${\mathfrak{u}}(n)$, the components of the 2-form curvature are
\begin{align}
F(\partial_\mu,\partial_\nu)\equiv F_{\mu\nu}=F_{\mu\nu}^AT^A=[{\tilde{{\cal{A}}}}_\mu,{\tilde{{\cal{A}}}}_\nu]_\star-i\Theta^{-1}_{\mu\nu}\bbbone_n \label{curv1-37}
\end{align} 
\begin{align}
F(\partial_\mu ,X_{(\rho\sigma)})\equiv F_{\mu(\rho\sigma)}=F^A_{\mu(\rho\sigma)}T^A=[{\tilde{{\cal{A}}}}_\mu,{\tilde{{\cal{A}}}}_{(\rho\sigma)}]_\star
-\Theta^{-1}_{\mu\rho}{{\tilde{\cal{A}}}}_\sigma -\Theta^{-1}_{\mu\sigma}{\tilde{{\cal{A}}}}_\rho \label{curv2-1-37}
\end{align} 
\begin{align}
F(X_{(\mu\nu)},X_{(\rho\sigma)})\equiv F_{(\mu\nu)(\rho\sigma)}=F^A_{(\mu\nu)(\rho\sigma)}T^A=\\ \nonumber
[{\tilde{{\cal{A}}}}_{(\mu\nu)},{\tilde{{\cal{A}}}}_{(\rho\sigma)}]_\star+\Theta^{-1}_{\rho\nu}{\tilde{{\cal{A}}}}_{(\mu\sigma)} + \Theta^{-1}_{\sigma\nu}{\tilde{{\cal{A}}}}_{(\mu\rho)}+\Theta^{-1}_{\rho\mu}{\tilde{{\cal{A}}}}_{(\nu\sigma)} +\Theta^{-1}_{\sigma\mu}{\tilde{{\cal{A}}}}_{(\nu\rho)}\label{curv3-1-37}
\end{align} 
\end{proposition}
\begin{proof}
The proof is similar to the one for proposition \ref{curvature37}.
\end{proof}
The corresponding $U(n,{\cal{M}})$-invariant action is then defined by
\begin{align}
S^\prime_{{\cal{G}}_2}({\tilde{A}}_\mu, {\tilde{{\cal{A}}}}_{(\mu\nu)})=-{{1}\over{\alpha^2}}\tr_{{\mathfrak{u}}(n)}
\int d^{2n}x\big(F_{\mu\nu}\star F_{\mu\nu}+F_{\mu(\rho\sigma)}\star F_{\mu(\rho\sigma)}+ F_{(\mu\nu)(\rho\sigma)}\star F_{(\mu\nu)(\rho\sigma)}\big)\label{eq:actionymna}
\end{align}
$S^\prime_{{\cal{G}}_2}({\tilde{A}}_\mu, {\tilde{{\cal{A}}}}_{(\mu\nu)})$ can be viewed as a $U(n)$ generalisation of \eqref{eq:actionym} in which the Higgs sector involves ${{1}\over{2}}D(D+1)$ copies of (the NC analog of) a Higgs pattern in the adjoint representation, as called in the physics litterature.\par 
We now compare the gauge invariant actions constructed within the present framework of derivation based differential calculus to other available gauge invariant noncommutative field theory models. \par 
Consider first the minimal NC differential calculus generated by ${\cal{G}}_1$. In the $U(1)$ case, where the module $\modM$ is equal to the Moyal algebra, the present framework gives rise (up to unessential rescaling) to the action $S_0\equiv S_{{\cal{G}}_1}={{1}\over{4}}\int d^Dx (F_{\mu\nu}\star F_{\mu\nu})$. We note that $S_{{\cal{G}}_1}$ is formally similar to the Connes-Lott action functional obtained from the non compact spectral triple proposed in \cite{GAYRAL2004} (for details on the corresponding construction see also \cite{MARSE1}, and \cite{MARSE2}). However, the (unital) algebra underlying this latter situation is a subalgebra of ${\cal{M}}$, basically the algebra of smooth bounded functions of $\mathbb{R}^{2n}$ having all their derivatives bounded. As far as field theory aspect is concerned, it is know that $S_{{\cal{G}}_1}$ has IR/UV mixing. The extension of the $U(n)$ case corresponding to the module $\modM=\algA^n$ is given by the action $S^\prime_{{\cal{G}}_1}$, \eqref{eq:unaction}. Notice that this action has also IR/UV mixing, stemming from the "$U(1)$ part" of the polarisation tensor which unfortunately cannot be disentangled from the whole $U(n)$ theory. The corresponding calculation is presented in the Appendix~\ref{uncasepolar}. \par 
Consider now the NC differential calculus based on ${\cal{G}}_2$. In the $U(1)$ case, we obtain the action \eqref{eq:actionym} $S_{{\cal{G}}_2}$ which supports a natural interpretation as a Yang-Mills-Higgs action on Moyal spaces. where Higgs fields are associated to the covariant coordinates in the additional directions defined by the derivations supplementing the spatial derivations. Note that the Higgs potential part of the action is rigidely fixed, being naturally identified to a component of the curvature. The action $S_{{\cal{G}}_2}$ together with its interpretation as a NC Yang-Mills-Higgs action on Moyal spaces within the present framework appear to be new. A similar comment applies to the action $S^\prime_{{\cal{G}}_2}$ \eqref{eq:actionymna} which provides a $U(n)$ generalisation 
of \eqref{eq:actionym}. In these gauge models defined by $S_{{\cal{G}}_2}$ or $S^\prime_{{\cal{G}}_2}$, the polarisation tensor for the gauge potential $A_\mu$ still exhibits an IR singularity similar to the one given in \eqref{eq:singul1} with however a different overall factor depending on the dimension $D=2n$ of the Moyal space and the Higgs fields content, which however does not vanish for any positive even integer values for $D$. The calculation is presented for a class of gauge models related to $S_{{\cal{G}}_2}$ in Appendix~\ref{vaccuum}  and can be easily performed by using auxiliary integrals given in this appendice. \par

\section{Discussion}\label{discussion}

The derivation-based differential calculus is a mathematical algebraic framework that permits one to generate from a given associative algebra different consistent differential calculi. The case of Moyal algebras has been considered in the present paper. Let us compare this latter situation with two other noncommutative geometries, which share some common structures with the one studied here.\par

First consider $M_n(\gC)$, the finite dimensional algebra of $n\times n$ matrices. The algebra $M_n(\gC)$ has only inner derivations, a trivial center $\gC$ and admits canonical NC gauge invariant connections. This last property is insured by the existence, for $\algA={\cal{M}}$ or $\algA=M_n(\gC)$, of a $\gC$-linear map $\eta: \der(\algA) = \Int(\algA) \rightarrow \algA$ such that $\kX(a) = [\eta(X), a]$ for any $a \in \algA$ ($\algA=M_n(\gC)$ or ${\cal{M}}$). This map defines a canonical gauge invariant connection on (the right $\algA$-module) $\algA$: $a \mapsto \nabla_{X} a = - a \eta(X)$. For the differential calculus based on a subalgebra of $\der({\cal{M}})$, as considered here, the map $\eta$ is defined by \eqref{eq:thetaform}. For the differential calculus based on $\der(M_n(\gC))$, the map is defined by the canonical $1$-form $i\theta$ of $M_n(\gC)$, $i\theta\in\Omega^1_{\der}(M_n(\gC))$ interpreted as a map $\Int(M_n(\gC))\to M_n(\gC)$ and such that $i\theta(\adrep_\gamma)=\gamma-{{1}\over{n}}\tr(\gamma){\mathbb{I}}$, $\forall\gamma\in M_n(\gC)$. However, this last map is an isomorphism of Lie algebras from $\der(M_n(\gC))$ to $ \ksl_n\subset M_n(\gC)$, the Lie algebra of traceless elements and therefore, the curvature of the canonical connection is zero. This is not the case for the differential calculus considered here. The map $\eta$ defined in \eqref{eq:thetaform} is not a Lie algebra morphism which is signaled by a non zero curvature for the canonical connection.\par 

Consider now the algebra $\algA=C^\infty(M)\otimes M_n(\gC)$ of matrix valued functions on a smooth finite dimensional manifold $M$ whose derivation-based differential calculus was first considered in \cite{Dubois-Violette:1989vq}.  In the present case, $\caZ(\algA) = C^\infty(M)$ and $\der(\algA) = [\der(C^\infty(M))\otimes \gone ] \oplus [ C^\infty(M) \otimes \der(M_n) ] = \Gamma(M) \oplus [C^\infty(M) \otimes \ksl_n]$ in the sense of Lie algebras and $C^\infty(M)$-modules. $\Gamma(M)$ is the Lie algebra of smooth vector fields on $M$. Then, for any derivation ${\cal{X}}\in\der(\algA)$, ${\cal{X}} = X + ad_\gamma$
with $X \in \Gamma(M)$ and $\gamma \in C^\infty(M) \otimes \ksl_n$, the traceless elements in $\algA$. Set $\algA_0=C^\infty(M) \otimes \ksl_n$. One can identify $\Int(\algA) =\algA_0 $ and $\out(\algA) = \Gamma(M)$. Therefore, one has both inner and outer derivations, contrary to what happens for ${\cal{M}}$. Finally, one has
$\Omega^\bullet_\der(\algA)= \Omega^\bullet(M) \otimes \Omega^\bullet_\der(M_n)$
with a differential $\hd = \dd + \dd'$, where $\dd$ is the de~Rham differential and 
$\dd'$ is the differential operating on the $M_n(\gC)$ part. The $1$-form related to the canonical connection is defined by $i\theta({\cal{X}})=\gamma$. As a map from $\der(\algA)$ to $\algA_0$, it defines a splitting of Lie algebras and $C^\infty(M)$-modules of the short exact sequence 
\begin{align}
0\longrightarrow\algA_0\longrightarrow\der(\algA)\longrightarrow\out(\algA)\longrightarrow0
\end{align}
while the map $\eta$ defined by \eqref{eq:thetaform} does not have a similar property. The canonical connection on (the right $\algA$-module) $\algA$ is defined from $-i\theta$ by $\nabla_{\cal{X}}(a)={\cal{X}}(a)-i\theta({\cal{X}})a=X(a)-a\gamma$, $\forall a\in\algA$ but is not gauge invariant while the corresponding curvature is zero, due to the above property of splitting of Lie algebras. Past classical studies of the corresponding gauge theories, with an action constructed mainly from the square of the curvature, gave rise to the intepretation of the gauge potential as involving two parts, one being the ``ordinary'' gauge theories and the other one identifiable as Higgs fields. Indeed, one can show that the simple action $\sim$ $\int d^Dx F_{\mu \nu} F^{\mu \nu}$ constructed using the corresponding curvature $F_{\mu \nu}$ exhibits non trivial vacuum states in the Higgs part, from which a mass generation on the ``ordinary'' gauge fields is a consequence. This situation has been generalized \cite{DBV-M, Masson:1999} to the case of the algebra of endomorphisms of a $SU(n)$-vector bundle in the sense that the situation of the trivial bundle correspond to the algebra of matrix-valued functions. Because of the possible non trivial global topology of the bundle, the situation is more involved \cite{DBV-M} but reveals essentially that this physical interpretation of the components of the noncommutative gauge fields can be performed in the same way. This framework has been used to generalize and classify NC generalisations of invariant connections in \cite{serie-masson}. \par

The Yang-Mills-Higgs type action constructed from differential calculus based on the subalgebra ${\cal{G}}_2\subset\der({\cal{M}})$ in Section~\ref{gaugemoyal} shares common features with this last situation: Each additional inner derivation supplementing the "ordinary spatial derivations", which may be viewed as related to an ``extra noncommutative dimension'', corresponds to an additional covariant coordinate that can be interpreted as a Higgs field. Covariant coordinates have thus a natural interpretation as Higgs fields within the framework of the present derivation-based differential calculus. Then, Yang-Mills-Higgs models can be obtained from actions built from the square of the curvature \eqref{curv1}-\eqref{curv3}. \par 

As a final remark, let us stress that until now, Higgs fields can be given a natural interpretation within NC geometry  when it is possible to introduce covariant coordinates as in Proposition~\ref{prop2.9}. In the particular examples explored so far, this occurs because it is possible to substract a ``reference object" to (some part of) the 1-form gauge connections, therefore canceling (some part of) the inhomogeneous term of gauge transformations. In the example considered in the present paper, this reference object is the $sp(2n,{\mathbb{R}})$ part of the 1-form $\eta$ appearing in Lemma~\ref{lemmainnerdifferential}. In the case of the NC geometry based on the algebra of matrix valued functions, the 1-form $i\theta$ in the matrix directions is used as the reference object. In both cases, the substraction takes place in the corresponding spaces of NC 1-forms. In the context of the standard model introduced using spectral triples (see e.g  \cite{Chamseddine:2006ep}), Higgs fields are introduced as covariant coordinates in the direction of the finite NC geometry, where the Dirac operator $D_F$ in that direction plays the role of the reference object. In this latter case, the substraction takes place in the space of bounded operators of the relevant Hilbert space because $D_F$ is simply constructed as a finite dimensional matrix and the connection 1-form is represented as bounded operators on the Hilbert space, through $d_Ua\mapsto[D,a]$ where $d_U$ is the differential for the universal differential calculus and $D$ is the (total) Dirac operator splitted into the spatial Dirac operator and $D_F$. We point out that the cancelation of some part of the inhomogeneous term of gauge transformations in all the above mentioned cases can be realized thanks to the existence of a relation between the differential in degree $0$ and the commutator with this reference object: in theses 3 cases, one has respectively $da=[\eta,a]$, $da=[i\theta,a]$ (for the matrix algebra) and $d_Ua\mapsto[D,a]$.\par

\vskip 1 true cm

\noindent
{\bf{Acknowledgements}}: We are grateful to M. Dubois-Violette for numerous discussions. This work has been supported by ANR grant NT05-3-43374 "GENOPHY". One of us (JCW) gratefully acknowledges partial support from the Austrian Federal Ministry of Science and Research, the High Energy Physics Institute of the Austrian Academy of Sciences and the Erwin Schr\"odinger International Institute of Mathematical Physics.\par

\appendix
\section{\texorpdfstring{The $U(n)$ case}{The U(n) case}.}\label{uncasepolar}
We denote by $T^A$ the generators of the ${\mathfrak{u}}(n)$ algebra, $A=0,1,...,n^2-1$, $T^0={{1}\over{{\sqrt{2n}}}}\bbbone_n$. The following relations hold : $[T^A,T^B]=if^{ABC}T^C$ and $\tr_{{\mathfrak{u}}(n)}(T^AT^B)={{1}\over{2}}\delta^{AB}$, $\forall A,B,C=0,1,...,n^2-1$, $\tr_{{\mathfrak{u}}(n)}(T^A)=0$, $\forall A=1,2,...,n^2-1$. The totally skew-symmetric structure constants $f^{ABC}$ verify $f^{0BC}=0$ and $\{T^A,T^B\}=d^{ABC}T^C$ where the symbol $d^{ABC}$ is totally symmetric.\par 
Useful relations are
\begin{equation}
f^{AMF}f^{BMF}=n\delta^{AB}(1-\delta^{A0});\quad d^{AMF}d^{BMF}=n\delta^{AB}(1+\delta^{A0});\quad f^{AMF}d^{BMF}=0 \label{eq:relatfd}
\end{equation}
In the $U(n)$ case, the Feyman rules used to compute the vacuum polarisation tensor for the simplest action 
$\int d^Dx\  \tr_{{\mathfrak{u}}(n)}(F_{\mu\nu}\star F_{\mu\nu})$ are
\begin{itemize}
\item 3-gauge boson vertex:
\begin{align}
(V^3)^{ABC}_{\mu\nu\rho}(p,q,r)=i(f^{ABC}\cos({{p\wedge q}\over{2}})+ d^{ABC}\sin({{p\wedge q}\over{2}}))\\ \nonumber
\big[(p-q)_\rho\delta_{\mu\nu}+(q-r)_\mu\delta_{\nu\rho}+(r-p)_\nu\delta_{\rho\mu}\big]
\end{align}
\item{Gauge boson-ghost vertex}
\begin{align}
(V^g)^{ABC}_{\mu}(p,q,r)=ip_{\mu}(f^{ABC}\cos({{p\wedge q}\over{2}})-d^{ABC}\sin({{p\wedge q}\over{2}}))
\end{align}
\item{4-gauge boson vertex (summation over $M$ is understood}
\begin{equation}
(V^4)^{ABCD}_{\mu\nu\rho\sigma}(p,q,r,s)=\nonumber
\end{equation}
\begin{align}
-\big[ (f^{ABM}\cos({{p\wedge q}\over{2}})+d^{ABM}\sin({{p\wedge q}\over{2}}))(f^{MCD}\cos({{r\wedge s}\over{2}})+d^{MCD}\sin({{r\wedge s}\over{2}}))(\delta_{\mu\rho}\delta_{\nu\sigma}-\delta_{\mu\sigma}\delta_{\nu\rho}) \nonumber
\\
+(f^{ACM}\cos({{p\wedge r}\over{2}})+d^{ACM}\sin({{p\wedge r}\over{2}}))(f^{MDB}\cos({{s\wedge q}\over{2}})+d^{MDB}\sin({{s\wedge q}\over{2}}))(\delta_{\mu\sigma}\delta_{\nu\rho}-\delta_{\mu\nu}\delta_{\sigma\rho}) \nonumber
\\
+(f^{ADM}\cos({{p\wedge s}\over{2}})+d^{ADM}\sin({{p\wedge s}\over{2}}))(f^{MBC}\cos({{q\wedge r}\over{2}})+d^{MBC}\sin({{q\wedge r}\over{2}}))(\delta_{\mu\nu}\delta_{\rho\sigma}-\delta_{\mu\rho}\delta_{\nu\sigma}) \nonumber
\\\big]
\end{align}
\end{itemize}
The diagrams contributing to the vacuum polarisation tensor are similar to the first three one depicted on the figure 1, with the expressions for the vertices given just above. The part contributing to the IR limit is given by
\begin{align}
(\omega_{\mu\nu}^{AB})^{IR}(p)=\int {{d^Dk}\over{(2\pi)^D}}{{1}\over{k^2(k-p)^2}}\big((V^3)^{AMN}_{\mu\alpha\beta}(p,-k,k-p)(V^3)^{AMN}_{\nu\alpha\beta}(-p,p-k,k)+\nonumber\\
(V^g)^{AMN}_{\mu}(-k,k-p,p)(V^g)^{AMN}_{\nu}(p-k,k,-p) \big)
\end{align}
By using \eqref{eq:relatfd}, $(\omega_{\mu\nu}^{AB})^{IR}(p)$ can be reexpressed as
\begin{equation}
(\omega_{\mu\nu}^{AB})^{IR}(p)=\int {{d^Dk}\over{(2\pi)^D}}{{1}\over{k^2(k-p)^2}}{\cal{P}}_{\mu\nu}(k,p)\delta^{AB}(2-\delta^{A0}\cos(k\wedge p))
\end{equation}
where ${\cal{P}}_{\mu\nu}(k,p)$ is a polynomial expression depending on $k$ and $p$. The second contribution only contributes to the IR/UV mixing and vanishes obviously when $A\ne0$, so that the ${\mathfrak{su}}(n)$ part of the vacuum polarisation tensor is free from IR/UV mixing.

\section{\texorpdfstring{Feynman rules for the $U(1)$ case.}{Feynman rules for the U(1) case.}}
\label{Feynman}
In the following vertex functions, momentum conservation is understood. All the momentum are entering. We define $p\wedge k$$\equiv$$p_\mu\Theta_{\mu\nu}k_\nu$.\par
\begin{itemize}
\item 3-gauge boson vertex:
\begin{align}
V^3_{\alpha\beta\gamma}(k_1,k_2,k_3)=-i2\sin({{k_1\wedge k_2}\over{2}})\big[(k_2-k_1)_\gamma\delta_{\alpha\beta}+(k_1-k_3)_\beta\delta_{\alpha\gamma}+(k_3-k_2)_\alpha\delta_{\beta\gamma}\big]
\end{align}
\item 4-gauge boson vertex:
\begin{align}
V^4_{\alpha\beta\gamma\delta}(k_1,k_2,k_3,k_4)=-4\big[ (\delta_{\alpha\gamma}\delta_{\beta\delta}-\delta_{\alpha\delta}\delta_{\beta\gamma})\sin({{k_1\wedge k_2}\over{2}})\sin({{k_3\wedge k_4}\over{2}}) \nonumber
\\
+(\delta_{\alpha\beta}\delta_{\gamma\delta}-\delta_{\alpha\gamma}\delta_{\beta\delta})\sin({{k_1\wedge k_4}\over{2}})\sin({{k_2\wedge k_3}\over{2}}) +(\delta_{\alpha\delta}\delta_{\beta\gamma}-\delta_{\alpha\beta}\delta_{\gamma\delta})\sin({{k_3\wedge k_1}\over{2}})\sin({{k_2\wedge k_4}\over{2}})\big]
\end{align}
\item gauge boson-ghost $V^g_{\mu}(k_1,k_2,k_3)$ and gauge boson-Higgs $V^H_{ab\mu}(k_1,k_2,k_3)$ vertices. \par 
We set $\Phi_a\equiv{\cal{A}}_{(\mu\nu)}$, $a=1,...,{{D(D+1)}\over{2}}$. Then:
\begin{align}
V^g_{\mu}(k_1,k_2,k_3)=i2k_{1\mu}\sin({{k_2\wedge k_3}\over{2}});\quad V^H_{ab\mu}(k_1,k_2,k_3)=i\delta_{ab}(k_1-k_2)_\mu\sin({{k_2\wedge k_3}\over{2}})
\end{align}
\item 
\begin{align}
V^s_{ab\alpha\beta}(k_1,k_2,k_3,k_4)=-2\delta_{\alpha\beta}\delta_{ab}\big[\cos({{k_3\wedge k_1+k_4\wedge k_2}\over{2}})-
\cos({{k_1\wedge k_2}\over{2}})\cos({{k_3\wedge k_4}\over{2}}) \big]
\end{align}
\item   3-Higgs $V^H_{abc}(k_1,k_2,k_3)$ vertex
\begin{align}
V^H_{abc}(k_1,k_2,k_3)=iC_{ab}^c\sin({{k_1\wedge k_2}\over{2}})
\end{align}
\item 4-Higgs vertex:
\begin{align}
V^H_{abcd}(k_1,k_2,k_3,k_4)=4\big[ (\delta_{ac}\delta_{bd}-\delta_{ad}\delta_{bc})\sin({{k_1\wedge k_2}\over{2}})\sin({{k_3\wedge k_4}\over{2}}) \nonumber
\\
+(\delta_{ab}\delta_{cd}-\delta_{ac}\delta_{bd})\sin({{k_1\wedge k_4}\over{2}})\sin({{k_2\wedge k_3}\over{2}}) +(\delta_{ad}\delta_{bc}-\delta_{ab}\delta_{cd})\sin({{k_3\wedge k_1}\over{2}})\sin({{k_2\wedge k_4}\over{2}})\big]
\end{align}
\end{itemize}

\section{\texorpdfstring{Computation of the vacuum polarisation for the $U(1)$ models.}{Computation of the vacuum polarisation for the U(1) models.}}\label{vaccuum}
The polarisation tensor for the gauge potential $A_\mu$ involved in $S_{{\cal{G}}_2}$ exhibits an IR singularity similar to the one given in \eqref{eq:singul1} with however a different overall factor depending on the dimension $D=2n$ of the Moyal space and the Higgs fields content. The calculation is easily performed by using auxiliary integrals given below.\par 
We consider here the case where no bare mass term for the gauge potential is present. Inclusion of a bare mass term would not alter the conclusion. We set $\Phi_a\equiv{\cal{A}}_{(\mu\nu)}$, $a=1,...,{{D(D+1)}\over{2}}$ and parametrize the gauge invariant action as
\begin{equation}
S_{cl}=\int d^Dx{{1}\over{4}}\big(F_{\mu\nu}\star F_{\mu\nu}+(D_\mu\Phi_a)^2+F_{ab}\star F_{ab}\big) \label{eq:actionsimple}
\end{equation}
where the coupling constant $\alpha$ has been set equal to $1$ and $D_\mu\Phi_a$$=$$[{\cal{A}}_\mu,\Phi_a]_\star$$=$$\partial_\mu\Phi_a-i[A_\mu,\Phi_a]_\star$. The action $S_{cl}$ must be supplemented by a BRST-invariant gauge fixing term $S_{GF}$, given by
\begin{equation}
S_{GF}=s\int d^Dx\big({\bar{C}}\partial^\mu A_\mu+{{\lambda}\over{2}}{\bar{C}}b \big)=\int d^Dx\big(b\partial^\mu A_\mu-{\bar{C}}\partial^\mu(\partial_\mu C-i[A_\mu,C]_\star)+{{\lambda}\over{2}}b^2\big)
\end{equation}
where the nilpotent Slavnov operation $s$ is defined by
\begin{equation}
sA_\mu=\partial_\mu C-i[A_\mu,C]_\star,\quad sC=iC\star C,\quad s{\bar{C}}=b,\quad sb=0
\end{equation}
where $\lambda$ is a real constant and $C$, ${\bar{C}}$ and $b$ denote respectively the ghost, the antighost and the Stuekelberg auxiliary field with ghost number respectively equal to $+1$, $-1$ and $0$. $s$ acts as a graded derivation on the various objects with grading defined by the sum of the degree of differential forms and ghost number (modulo 2). In the following, we ujse a Feynman-type gauge. Accordingly, the propagator for the $A_\mu$ in momentum space takes the diagonal form 
$G_{\mu\nu}(p)$$=$$\delta_{\mu\nu}/p^2$. The ghost and Higgs propagators are respectively given by $G_g(p)$$=$$1/p^2$ and $G^H_{ab}(p)$$=$$2\delta_{ab}/(p^2+\mu^2)$. The relevant Feynman rules are given in the Appendix~\ref{Feynman}.\par
The one-loop diagrams contributing to the vacuum polarisation tensor $\omega_{\mu\nu}(p)$ are depicted on the figure 1. The respective contributions can be written as
\begin{align}
\omega^1_{\mu\nu}(p)=4\int {{d^Dk}\over{(2\pi)^D}}{{\sin^2({{p\wedge k}\over{2}})}\over{k^2(p+k)^2}}\big[((k-p)^2+(k+2p)^2)\delta_{\mu\nu}+(D-6)p_\mu p_\nu
\\
+(p_\mu k_\nu+k_\mu p_\nu)(2D-3)+k_\mu k_\nu(4D-6)\big]\label{eq:omega1}
\end{align}
\begin{align}
\omega^2_{\mu\nu}(p)=4\int {{d^Dk}\over{(2\pi)^D}} {{\sin^2({{p\wedge k}\over{2}})}\over{k^2(p+k)^2}}k_\mu k_\nu;\quad
\omega^3_{\mu\nu}(p)=8(D-1)\delta_{\mu\nu}\int {{d^Dk}\over{(2\pi)^D}} {{\sin^2({{p\wedge k}\over{2}})}\over{k^2}}\label{eq:omega23}
\end{align}
\begin{align}
\omega^4_{\mu\nu}(p)=4{\cal{N}}\int {{d^Dk}\over{(2\pi)^D}} {{\sin^2({{p\wedge k}\over{2}})}\over{(k^2+\mu^2)((p+k)^2+\mu^2)}}(p+2k)_\mu(p+2k)_\nu \label{eq:omega4}
\end{align}
\begin{align}
\omega^5_{\mu\nu}(p)=-4{\cal{N}}\delta_{\mu\nu}\int {{d^Dk}\over{(2\pi)^D}} {{\sin^2({{p\wedge k}\over{2}})}\over{(k^2+\mu^2)}} \label{eq:omega5}
\end{align}
where ${\cal{N}}$ is the number of $\Phi$ fields, i.e ${\cal{N}}$$=$${{D(D+1)}\over{2}}$ for $sp(D,{\mathbb{R}})$. \par 
\begin{figure}[!htb]
  \centering
  \includegraphics[scale=0.4]{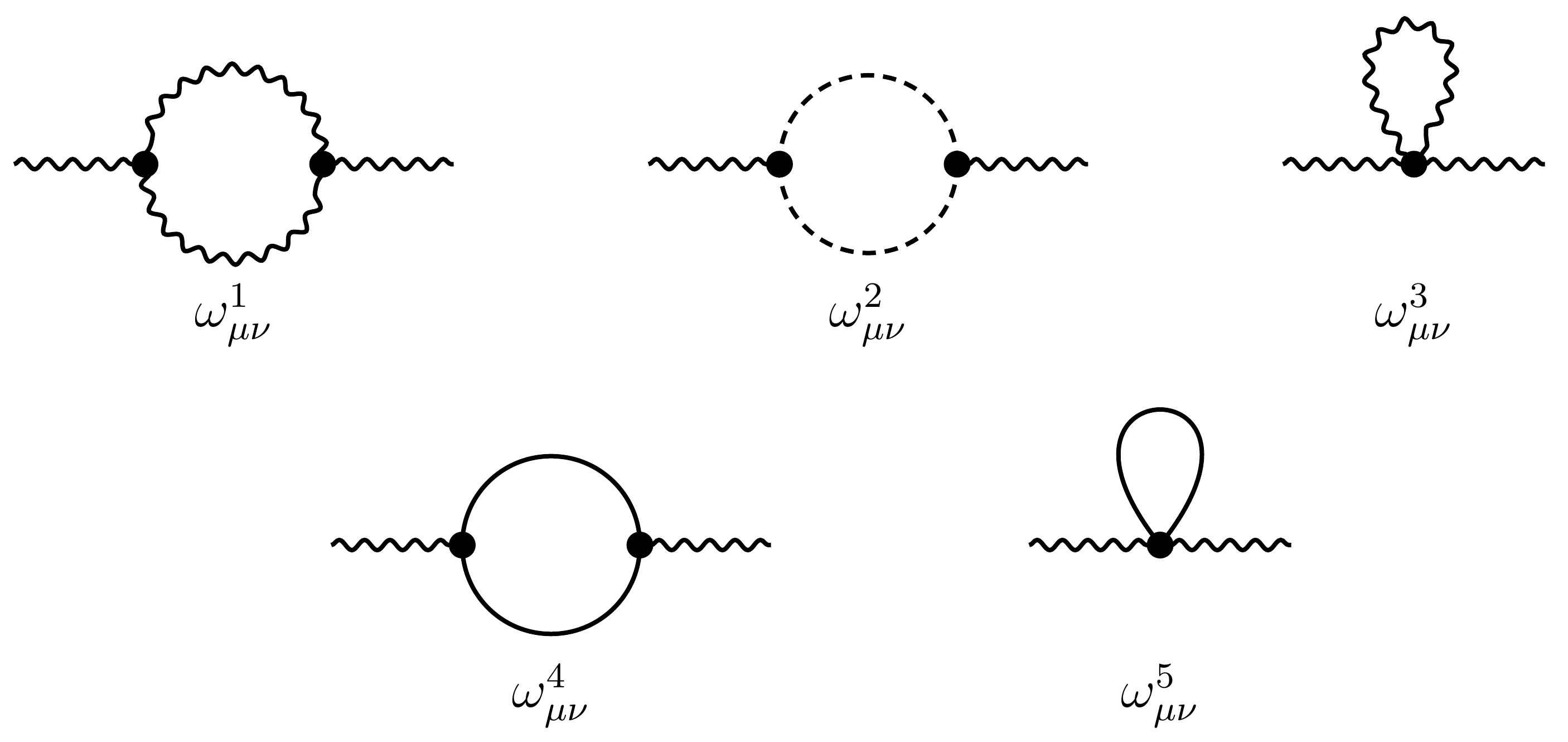}
  \caption[]{\footnotesize{One-loop diagrams contributing to the vacuum polarisation tensor. The wavy lines correspond to $A_\mu$. 
The full (resp. dashed) lines correspond to the $\Phi$ (resp. ghost) fields}}
  \label{fig:figure1}
\end{figure}
By using ${{1}\over{ab}}$$=$$\int_0^1 dx{{1}\over{(xa+(1-x)b)^2}}$ and standard manipulations, one extracts the IR limit of \eqref{eq:omega1}-\eqref{eq:omega5}, denoted by $\omega^{IR}_{\mu\nu}(p)$. In the course of the derivation, we further use:
\begin{equation}
J_N({\tilde{p}})\equiv\int {{d^Dk}\over{(2\pi)^D}}{{e^{ik{\tilde{p}}}}\over{(k^2+m^2)^N}}=a_{N,D}{\cal{M}}_{N-{{D}\over{2}}}(m|{\tilde{p}}|)
\label{eq:JN}
\end{equation}
\begin{equation}
J_{N,\mu\nu}({\tilde{p}})\equiv\int {{d^Dk}\over{(2\pi)^D}}{{k_\mu k_\nu  e^{ik{\tilde{p}}}}\over{(k^2+m^2)^N}}=a_{N,D}\big(\delta_{\mu\nu}{\cal{M}}_{N-1-{{D}\over{2}}}(m|{\tilde{p}}|)-{\tilde{p}}_\mu{\tilde{p}}_\nu{\cal{M}}_{N-2-{{D}\over{2}}}(m|{\tilde{p}}|) \big) \label{eq:JMU}
\end{equation}
where
\begin{equation}
a_{N,D}={{2^{-({{D}\over{2}}+N-1)}}\over{\Gamma(N)\pi^{{D}\over{2}}}};\quad {\cal{M}}_Q(m|{\tilde{p}}|)={{1}\over{(m^2)^Q}}(m|{\tilde{p}}|)^Q{\bf{K}}_Q(m|{\tilde{p}}|)
\end{equation}
in which ${\bf{K}}_Q(z)$ is the modifed Bessel function of order $Q$$\in$${\mathbb{Z}}$ (recall ${\bf{K}}_{-Q}(z)$$=$${\bf{K}}_Q(z)$) together with the asymptotic expansion
\begin{equation}
{\cal{M}}_{-Q}(m|{\tilde{p}}|)\sim2^{Q-1}{{\Gamma(Q)}\over{{\tilde{p}}^{2Q}}},\quad Q>0
\end{equation}
The IR limit of the vacuum polarisation tensor is given by
\begin{equation}
\omega^{IR}_{\mu\nu}(p)=(D+{\cal{N}}-2)\Gamma({{D}\over{2}}) {{{\tilde{p}}_\mu{\tilde{p}}_\nu}\over{{\pi^{D/2}(\tilde{p}}^2)^{D/2}}}+...
\end{equation}
where the ellipses denote subleading singular terms. The overall factor affecting $\omega_{\mu\nu}^{IR}(p)$ is modified compared to \eqref{eq:singul1} but cannot be canceled by tuning the values for $D$ and ${\cal{N}}$. \par 

A similar calculation using \eqref{eq:JN} and \eqref{eq:JMU} leads to the expression of the IR limit of the two-point function for the Higgs fields given by
\begin{align}
\Pi_{ab}^{IR}(p)={{1}\over{4\pi^{D/2}}}[D+5-4{\cal{N}}]\delta_{ab}{{\Gamma({{D}\over{2}}-1)}\over{{\tilde{p}}^{2({{D}\over{2}}-1)}}}
\end{align}
\begin{figure}[!htb]
  \centering
  \includegraphics[scale=1]{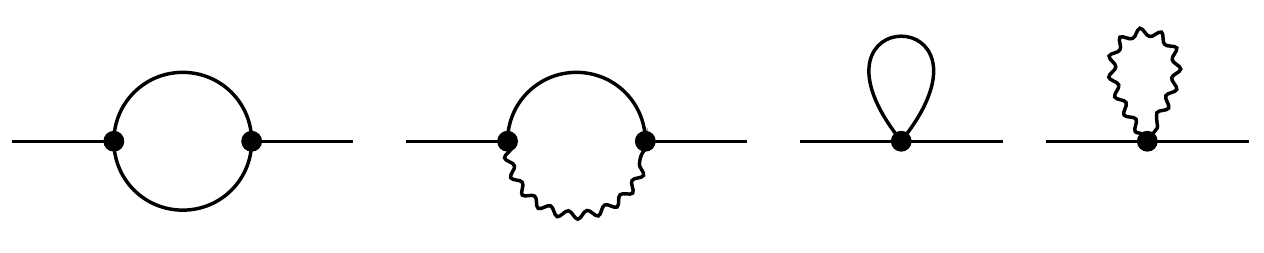}
  \caption[]{\footnotesize{One-loop diagrams contributing to the vacuum polarisation tensor. The wavy lines correspond to $A_\mu$. 
The full (resp. dashed) lines correspond to the $\Phi$ (resp. ghost) fields}}
  \label{fig:figure1}
\end{figure}
As expected, the overall factor affecting $\omega_{\mu\nu}^{IR}(p)$ is modified compared to \eqref{eq:singul1} but cannot be canceled by tuning the values for $D$ and ${\cal{N}}$. Starting from a subalgebra of the full derivation algebra $isp(D,{\mathbb{R}})$ would not have modified these conclusions.

\end{document}